\documentclass[article]{article}

\usepackage{hyperref}
\usepackage{graphicx}
\usepackage{color}
\usepackage{amsmath,amssymb}
\usepackage{bm}
\usepackage{mathrsfs}
\usepackage[margin=1in]{geometry}
\usepackage[authoryear]{natbib}
\usepackage{caption}
\usepackage{subcaption}

\begin{document}


\title{Combining mechanical and chemical effects in the deformation and failure of a cylindrical electrode particle in a Li-ion battery}
\author{Jeevanjyoti Chakraborty\footnote{Corresponding author. Tel: +44-1865-615144; E-mail address: \texttt{jeevanjyoti4@gmail.com}, \texttt{chakraborty@maths.ox.ac.uk}}, Colin P. Please, Alain Goriely, S. Jonathan Chapman \\ Mathematical Institute, University of Oxford, Oxford, OX2 6GG, UK}

\maketitle

\begin{abstract}
A general framework  to study the mechanical behaviour of a cylindrical silicon anode particle in a lithium ion battery as it undergoes lithiation is presented. The two-way coupling between stress and concentration of lithium in silicon, including the possibility of plastic deformation, is taken into account and two particular cases are considered. First,  the cylindrical particle is assumed to be free of surface traction  and  second, the axial deformation of the cylinder is prevented. In both  cases plastic stretches develop through the entire cylinder and not just near the surface as is commonly found in spherical anode particles. It is  shown that the stress evolution depends both on the  lithiation rate and the external constraints.  Furthermore, as the cylinder expands during lithiation it can develop a compressive axial stress large enough to induce buckling, which in turn may lead to mechanical failure. An explicit criterion for swelling-induced buckling  obtained as a modification of the classical Euler buckling criterion shows the competition between the stabilising effect of radius increase and the destabilising effect of axial stress.
\end{abstract}

\section{Introduction}

The lithium-ion battery has become the forerunning energy storage medium for numerous electronic devices including laptop computers and mobile phones \citep{2001NatureTarasconArmand, 2011BookFletcher, 2012ProcIEEEWhittingham}. The portability of these devices requires that their power sources be both light-weight and that they have large storage capacity. These requirements are met in the lithium ion battery as lithium (Li) is the lightest metal, and lithium-ion batteries have a high energy density. However, other technological challenges are still present \citep{2012AngewRevChoi, 2013JACSGoodenough}. The biggest of these is to make lithium-ion batteries viable for use in electric transportation systems at large scales \citep{2011EnergyEnvironSciScrosati, 2012EnergyEnvironSciThackeray}, so that dependence on fossil fuels may be reduced. This requires  lithium-ion batteries with even higher energy densities. Since energy density is fundamentally linked to the intrinsic chemistry of the battery materials, this requirement necessitates the consideration of new materials beyond the traditional ones: most commonly, graphite for the anode, and cobalt-oxide, manganese oxide or iron phosphate for the cathodes. Possibly the best candidate for this new material -- at least for the anode -- is silicon (Si). In the fully lithiated state, up to 4.4 atoms of Li may be accommodated for every atom of Si giving for the equilibrium (amorphous) phase, Li$_{22}$Si$_5$ \citep{1964KrisGlad}. This is far more Li-rich than graphite, which gives LiC$_6$ in the fully lithiated state. The result is a theoretical specific capacity value of 4200 mAhg$^{-1}$ for Si compared to  372 mAhg$^{-1}$ for graphite. This higher specific capacity leads, in turn, to a higher energy density. 

Unfortunately -- and, perhaps manifesting the ``law of conservation of difficulty" -- it is this very ability of Si to accommodate high amounts of Li which presents our biggest challenge. Such high intake of lithium is accompanied by an extreme volume change in Si which can be as high as 310\% for the fully lithiated state \citep{2001ECMSSLBeaulieu}. This volume change induces stress within the silicon due to both spatially non-uniform Li concentration and externally imposed geometrical constraints. Under conditions of cyclic charging/discharging these stresses can lead to mechanical fracture of the Si anode particles, and ultimately to a decay of the specific capacity \citep{2007JPowerSourcesKasavajjula, 2008NatureNanotechCui, 2013ScienceEbner}. Since for a given charging rate, smaller Si particles will have a lower spatial inhomogeneity in Li concentration, this problem may be circumvented to a considerable extent by using nano-structured Si anode particles \citep{2008NatureNanotechCui, 2009NanoLettCui, 2010NanoLettSong, 2011ScienceKovalenko}; see also the recent reviews by \citet{2013JMaterChemAZamfir}, \citet{2014JPhysChemLettSong}, and \citet{2014AdvEnergyMaterSu}, and references therein. 

In a battery electrode, the Si nanoparticles are geometrically constrained due to physical contact with other particles and supporting substrates, and their relative expansion leads to stress and possible mechanical failure. Common forms of nanoparticle are Si wires and pillars; we therefore consider here an idealized circular cylindrical anode particle. Such particles have an additional failure mechanism over spherical particles in that they may buckle under sufficiently large axial loads through an Euler buckling instability \citep{govade08}. Such buckling may lead to a deterioration of the battery performance. Indeed, the significant increase in length that may occur in Si nanowires during lithiation was clearly shown in the pioneering work by \citet{2008NatureNanotechCui}. They established that compared to the thin film or spherical particle geometry, the cylindrical geometry of the nanowire is far less prone to pulverization even though it undergoes increase in diameter and length. They also showed that when constrained against free axial growth by a rigid backbone (in the form of a thin Ni coating), the nanowire buckled into a helical shape; see Fig. 3 in the paper by \citet{2008NatureNanotechCui}. Later, the susceptibilty of Si nanowires to buckling (for sufficiently slender geometries) was confirmed through molecular dynamics simulations \citep{2012CompositesBMDS}. Interestingly, buckled shapes of Si nanowires can also be clearly seen in a study by \citet{2013ACSNanoLiu} (see Fig. 2 (g)-(k) in their paper) even though the focus of their study was on self-limiting lithiation. Despite such clear evidence, so far, there seems to be very little discussion from a modelling perspective on the entire aspect of buckling related to battery electrode particles. The notable exception is the work by \citet{2012JMechPhysSolidsBhandakkar} where, however, their interest is at the electrode level (having a honeycomb architecture). A clear framework focussing on length increase of individual electrode particles and its consequences on buckling when subject to constraints (as they inevitably will be in an actual battery configuration) seems to be lacking in the literature. Accordingly, in this paper, we present a theoretical basis which may provide design criteria for fabricating electrode particles accounting for buckling failure.

We formulate a generic framework to consider the mechanical behaviour of a cylindrical Si anode particle as it undergoes lithiation. Importantly, we incorporate the two-way coupling between stress and concentration of Li in Si; physically, this means that we account for the effects of both diffusion-induced stresses as well stress-enhanced diffusion. A number of recent works \citep{2010JElectrochemSocSethuraman, 2011JMechPhysSolidsBower, 2011JAmCeramSocZhao, 2011JAPGaoZhou, 2011JPowerSourcesHaftbaradaran, 2012JMechPhysSolidsAnand, 2012ActaMechSinicaGao, 2013SciRepLevitas, 2013JPhysDSong, 2014JMechPhysSolidsBucci, 2014JAMGuo} have considered similar problems, all based on the seminal work by \citet{1973ActaMetallLarcheCahn}. Our framework, however, is based on the recent work by \citet{2012JMechPhysSolidsCui} who presented a new stress-dependent chemical potential for the finite deformation of solids building on the contribution by \citet{2001JMechPhysSolidsWu}; indeed, their chemical potential reduces to those of \citet{2001JMechPhysSolidsWu} and of \citet{1973ActaMetallLarcheCahn} as special cases. Furthermore, as an important addition to previous works on cylindrical Si particles \citep{2011JMechPhysSolidsRyu, 2011JAPGaoZhou, 2014JAPZhang}, we incorporate the possibility of plastic deformations in our framework. This allows us to probe situations with relatively high charging rates which induce stresses high enough to reach the yield strength of Si. Furthermore, the absence of symmetry along two directions present in the previously studied thin film \citet{2011JMechPhysSolidsBower, 2014JMechPhysSolidsBucci} and spherical \citet{2012JMechPhysSolidsCui} cases necessitates a careful formulation of the chemo-mechanical model along three directions (radial, circumferential, and axial). We then use this generic framework to study the buckling limits of the cylindrical particle, examining in detail the competitive roles played by the growing stresses and the growing radius.

The remainder of the paper is organized as follows. In Sec.~\ref{sec:formulation}, we present the aforementioned formulation for the general case (Sec.~\ref{subsec:general}) considering the deformation of a cylindrical Si anode particle undergoing lithiation. We then discuss a situation (Sec.~\ref{subsec:special}), where the Si anode particle is constrained in the axial direction. We also present a framework to determine the critical buckling lengths based on the classical Euler buckling formula but with two important modifications: first, incorporating the effect of the changing radius with time; and the second, incorporating the effect of the changing modulus of elasticity with concentration. In Sec.~\ref{sec:resdisc}, we present the results corresponding to the axially unconstrained case, that is when all surfaces of the cylinder are free from external physical constraint (Sec.~\ref{subsec:generalresults}) and the axially constrained case (Sec.~\ref{subsec:specialresults}) discussing, in detail, the evolution of the stresses and the plastic stretches with physical interpretations, and highlighting the important differences between the two cases as well those with the spherical case of \cite{2012JMechPhysSolidsCui}. We then examine in detail the buckling limits using the modified criteria. 

\section{Mathematical Formulation} \label{sec:formulation}

We consider the model problem of a single cylindrical-shaped silicon anode particle which undergoes deformation due to the charging (or, discharging) process as lithium atoms are inserted into (or, taken out of) the silicon. We assume that the charging (or, discharging) takes place uniformly all around the periphery, and that the entire process is axisymmetric. We present a general model for axisymmetric deformation and consider the two cases in which the cylinder is (i) unconstrained and (ii) physically constrained against deformation in the axial direction. The latter situation provides a natural setting to investigate the possibility of mechanical ``failure" of the electrode through buckling, and we present a framework to study such behaviour.  

\subsection{General Case} \label{subsec:general}

\subsubsection{Deformation gradient decomposition} 

Lithiation changes the electrode from a reference configuration $\mathscr{B}$ to a deformed configuration $\mathscr{B}^\prime$ via a smooth, one-to-one function $\varphi$, referred to as the deformation map which maps a point $\bm{X} \in \mathscr{B}$ to a point $\bm{x} \in \mathscr{B}^\prime$. The difference between the final location $\bm{x}=\varphi(\bm{X})$ of a material point and the initial location $\bm{X}$ is the displacement field
\begin{equation}
\bm{u} (\bm{X}) = \varphi(\bm{X}) - \bm{X}.
\end{equation} 
We use the standard cylindrical coordinates $(r,\theta,z)$ in $\mathscr{B}$ 
so that a point in the reference configuration is $\mathbf{X}=r \mathbf{e}_r+z \mathbf{e}_z$.
The displacement field vector is made up of three components $u$, $v$, and $w$ along the radial, azimuthal, and axial directions, such that $\bm{u}= u\mathbf{e}_r+v\mathbf{e}_\theta+w\mathbf{e}_z\equiv [u \; v \; w]^T$. The azimuthal direction will be also referred to as the hoop direction. Our assumption of axial symmetry implies that $v \equiv 0$.

It is important to note that the deformation $\varphi(\bm{X})$ is a cumulative manifestation of various mechano-chemical processes occurring during lithiation or delithiation; hence, the displacement field, $\bm{u} (\bm{X})$ is not limited to elastic deformations only. To delineate and model the contributions of these processes, the usual practice is to decompose the total deformation into three parts: an elastic part which is reversible, an expansion part corresponding to the volume change as a result of lithiation, and an irreversible plastic part which allows no change in volume. Such a decomposition may be carried out mathematically in terms of the strain, the total stretch, or the total deformation gradient. The third approach of using the total deformation gradient, first proposed by \citet{1969JAMLee}, is particularly suitable for Li/Si electrodes. In this approach, the deformation gradient, defined as $\mathbf{F}(\bm{X}) = \nabla \varphi(\bm{X})$, is decomposed as
\begin{equation}
\mathbf{F} = \mathbf{F}^e \mathbf{F}^{\rm{SF}} \mathbf{F}^p, \label{eq:Fdecomp}
\end{equation}
where $\mathbf{F}^e$, $\mathbf{F}^{\rm{SF}}$, and $\mathbf{F}^p$ are the deformation gradients corresponding to the elastic, stress-free volumetric, and plastic deformation parts respectively. Note that the order of these deformation gradient constituents do not matter in Eq.~\ref{eq:Fdecomp} in the absence of anisotropy, which as explicated in the following is indeed the case we are interested in here. A physically intuitive explanation in the context of batteries of this decomposition is presented by \citet{2013JMechSciTechReview}. Interestingly, this kind of decomposition has been used -- albeit, without the plastic component -- in a number of different settings: for instance, in tumour growth by Ambrosi and co-workers \citet{2002IJEngSciAmbrosiMollica, 2007MathMechSolidsAmbrosiGuana}, more recently by \citep{2012RoyalMaclaurinChapman}; and in various morphoelasticity models by Goriely and co-workers \citep{2005PRLGorielyBenAmar, 2005JMPSBenAmarGoriely, 2009JBioDynVandiverGoriely, 2011JMPSMoultonGoriely, 2013JMPSMoultonLessinnesGoriely}, following early work by \citet{1994JBiomechRodriguez}. In this decomposition, the deformation gradient $\mathbf{F}^{\rm{SF}}$ is referred to as ``stress-free" because it is due solely to the unconstrained - hence, free of stress - shape change associated with lithium insertion or extraction. For simplicity we take this volumetric change to be isotropic in nature so that 
\begin{equation}
\mathbf{F}^{\rm{SF}} = (J^c)^{1/3} \mathbf{I},
\end{equation}
where $J^c=1+3\eta x_{\rm{max}}c$ with $\eta$ being the coefficient of compositional expansion, $x_{\rm{max}}=4.4$ the maximum value of $x$ (the number of moles of Li per mole of Si), which denotes the saturation of Li in Li$_x$Si, and $c=x/x_{\rm{max}}$ representing a non-dimensional measure of the Li concentration. Note that since the $\mathbf{F}^{\rm{SF}}$ is a multiple of the identity, the volumetric change will not create any residual stress in the material when the concentration $c$ is uniform. For the axisymmetric configuration we are considering the deformation gradient tensors $\mathbf{F}^e$ and $\mathbf{F}^p$ will also be diagonal, so that there are no shear stresses in the material. 

Since the deformation gradient $\mathbf{F}^p$ associated with the plastic part conserves volume we have $\det(\mathbf{F}^p) = 1$. Then, denoting by $\lambda_r$, $\lambda_\theta$, and $\lambda_z$ the plastic stretches in the radial, azimuthal, and axial directions, respectively, we have, in the standard representation of the deformation gradient in cylindrical coordinates,
\begin{gather}
\det(\mathbf{F}^p) = \det \left( \mathrm{diag}\left(
		\lambda_r, \lambda_\theta , \lambda_z\right )
	\right) =  \lambda_r \lambda_\theta \lambda_z = 1,
\end{gather}
Note that the simplification appearing in the works by \citet{2011JMechPhysSolidsBower} and \citet{2012JMechPhysSolidsCui},  in which two directions are indistinguishable (so that the plastic stretches are equal), is not possible in the present cylindrical case. 

The total deformation gradient may be written as
\begin{equation}
\mathbf{F} = \mathrm{diag}\left(
		1+\frac{\partial u}{\partial r},
	1+\frac{u}{r} ,1+\frac{\partial w}{\partial z} \right),
\end{equation}
and we define $J=\mathrm{det}(\mathbf{F}))$. The elastic part of the deformation gradient then becomes
\begin{equation}
\mathbf{F}^e = \mathbf{F} (\mathbf{F}^p)^{-1} (\mathbf{F}^{\rm{SF}})^{-1} 
	     = (J^c)^{-1/3} \mathrm{diag}\left(
				\frac{1+\partial u/\partial r}{\lambda_r} , \frac{1+u/r}{\lambda_\theta} , \frac{1+\partial w /\partial z}{\lambda_z} \right). 	
\end{equation}
This gives the elastic strain as
\begin{equation}
\mathbf{E}^e = \frac{1}{2} \left[ (\mathbf{F}^e)^T \mathbf{F}^e - \mathbf{I} \right] =\mathrm{diag}\left(
											E_r^e , E_\theta^e , E_z^e  \right),
\end{equation}
where, 
\begin{subequations} \label{eq:dim_strain-displacement}
\begin{align}
E_r^e &= \frac{1}{2} \left[ \left( F_r^e \right)^2 - 1 \right] = \frac{1}{2} \left(J^c\right)^{-2/3} \frac{\left( 1+\partial u/\partial r \right)^2}{\lambda_r^2} - \frac{1}{2}, \\
E_\theta^e &= \frac{1}{2} \left[ \left( F_\theta^e \right)^2 - 1 \right] = \frac{1}{2} \left(J^c\right)^{-2/3} \frac{\left( 1+u/r \right)^2}{\lambda_\theta^2} - \frac{1}{2}, \\
E_z^e &= \frac{1}{2} \left[ \left( F_z^e \right)^2 - 1 \right] = \frac{1}{2} \left(J^c\right)^{-2/3} \frac{\left( 1+\partial w/\partial z \right)^2}{\lambda_z^2} - \frac{1}{2}.
\end{align}
\end{subequations}
\subsubsection{Elastic deformation}
We assume that the elastic strains remain small during the deformation. Therefore, we use a strain-energy density function  in the reference frame of the form
\begin{align}
W = \frac{J^c}{2} \frac{E(c)}{1+\nu} \left[ (1-\nu)\left(E_{kk}^e\right)^2 + E_{jk}^e E_{kj}^e\right],
\end{align}
where $\nu$ is Poisson's ratio and $E(c)$ is a concentration dependent Young's modulus (see below). From the strain-energy density the non-zero components of the first Piola-Kirchhoff stress, $\mathbf{P} = \mathrm{diag}(\sigma_r^0, \sigma_\theta^0, \sigma_z^0)$, may be determined as
\begin{subequations} \label{eq:dim_PK-strains}
\begin{align}
\sigma_r^0 &= \frac{1}{F_r^*} \frac{\partial W}{\partial F_r^e} = J^c \frac{E(c)}{(1+\nu)(1-2\nu)} \left[ (1-\nu)E_r^e + \nu \left( E_\theta^e + E_z^e  \right)  \right] \frac{2E_r^e + 1}{1+\partial u/\partial r},\\ 
\sigma_\theta^0 &= \frac{1}{F_\theta^*} \frac{\partial W}{\partial F_\theta^e} = J^c \frac{E(c)}{(1+\nu)(1-2\nu)} \left[ (1-\nu)E_\theta^e + \nu \left( E_z^e + E_r^e  \right)  \right] \frac{2E_\theta^e + 1}{1+ u/r}, \\
\sigma_z^0 &= \frac{1}{F_z^*} \frac{\partial W}{\partial F_z^e} = J^c \frac{E(c)}{(1+\nu)(1-2\nu)} \left[ (1-\nu)E_z^e + \nu \left( E_r^e + E_\theta^e  \right)  \right] \frac{2E_z^e + 1}{1+\partial w/\partial z}.
\end{align}
\end{subequations}
The condition for mechanical equilibrium Div$\mathbf{P}=0$ leads to a single equation
\begin{equation}
\frac{\partial \sigma_r^0}{\partial r} + \frac{\sigma_r^0 - \sigma_\theta^0}{r} = 0. \label{eq:mecheq}
\end{equation}
The Cauchy stresses $\bm{\sigma}=\mathrm{diag}(\sigma_r , \sigma_\theta , \sigma_z)$ are related to the Piola-Kirchhoff stress by $\bm{\sigma}=J^{-1} \mathbf{P} \mathbf{F}^T$ so that 

\begin{subequations} \label{eq:dim_Cauchy}
\begin{align}
\sigma_r &= \frac{E}{(1+\nu)(1-2\nu)}\left[ (1-\nu)E_r^e + \nu\left( E_\theta^e + E_z^e  \right)  \right] \frac{\sqrt{2E_r^e + 1}}{\sqrt{2E_\theta^e + 1}\sqrt{2E_z^e + 1}}, \\
\sigma_\theta &= \frac{E}{(1+\nu)(1-2\nu)}\left[ (1-\nu)E_\theta^e + \nu\left( E_z^e + E_r^e  \right)  \right] \frac{\sqrt{2E_\theta^e + 1}}{\sqrt{2E_z^e + 1}\sqrt{2E_r^e + 1}}, \\
\sigma_z &= \frac{E}{(1+\nu)(1-2\nu)}\left[ (1-\nu)E_z^e + \nu\left( E_r^e + E_\theta^e  \right)  \right] \frac{\sqrt{2E_z^e + 1}}{\sqrt{2E_r^e + 1}\sqrt{2E_\theta^e + 1}}.
\end{align}
\end{subequations}
\subsubsection{Plastic flow}
The rate of the plastic deformation gradient is
\begin{equation}
\mathbf{D}^p = \dot{\mathbf{F}}^p (\mathbf{F}^p)^{-1} = \mathrm{diag}\left(
								\frac{\dot{\lambda}_r}{\lambda_r} , \frac{\dot{\lambda}_\theta}{\lambda_\theta} ,\frac{\dot{\lambda}_z}{\lambda_z} \right), \label{eq:rate_plastic}
\end{equation}
where a dot indicates the time derivative in the reference frame. We take the constitutive equation describing the viscoplastic behaviour of lithiated silicon as
\begin{equation}
\mathbf{D}^p = \frac{\partial G(\sigma_{\rm{eff}})}{\partial \bm{\tau}}, \label{eq:const}
\end{equation}
where $\bm{\tau} = \bm{\sigma} - \frac{1}{3}\rm{tr}(\bm{\sigma})\mathbf{I}$
is the deviatoric part of the Cauchy stress, the effective stress
\begin{gather}
\sigma_{\rm{eff}} = \sqrt{\frac{3}{2}} \sqrt{\bm{\tau}:\bm{\tau}} = \sqrt{\frac{3}{2}} \sqrt{\tau_r^2 + \tau_\theta^2 + \tau_z^2} \label{eq:dim_effectivestress}
\end{gather}
and $G$ is the flow potential. We adopt the same flow potential as \citet{2012JMechPhysSolidsCui}, namely
\begin{equation}
G(\sigma_{\rm{eff}}) = \frac{\sigma_f \dot{d}_0}{m+1} \left( \frac{\sigma_{\rm{eff}}}{\sigma_f} - 1  \right)^{m+1} H \!\! \left( \frac{\sigma_{\rm{eff}}}{\sigma_f} - 1 \right), \label{eq:flow_pot}
\end{equation}
where $\sigma_f$ is the initial yield stress of Si, $\dot{d}_0$ is the characteristic strain rate for plastic flow in Si, $m$ is the stress exponent for plastic flow in Si, and $H$ is the Heaviside step function. This is slightly different to that used by \citet{2011JMechPhysSolidsBower}, and was adapted to ensure a smooth transition from the elastic to the plastic regimes. Thus we obtain three equations for the three unknown plastic stretches
\begin{align}
\frac{\dot{\lambda}_r}{\lambda_r} = \sqrt{\frac{3}{2}} \dot{d}_0 \left( \frac{\sigma_{\rm{eff}}}{\sigma_f} - 1 \right)^m \frac{\tau_r}{\sqrt{\tau_r^2 + \tau_\theta^2 + \tau_z^2}} \, H \!\! \left( \frac{\sigma_{\rm{eff}}}{\sigma_f} - 1 \right), \\
\frac{\dot{\lambda}_\theta}{\lambda_\theta} = \sqrt{\frac{3}{2}} \dot{d}_0 \left( \frac{\sigma_{\rm{eff}}}{\sigma_f} - 1 \right)^m \frac{\tau_\theta}{\sqrt{\tau_r^2 + \tau_\theta^2 + \tau_z^2}} \, H \!\! \left( \frac{\sigma_{\rm{eff}}}{\sigma_f} - 1 \right), \\
\frac{\dot{\lambda}_z}{\lambda_z} = \sqrt{\frac{3}{2}} \dot{d}_0 \left( \frac{\sigma_{\rm{eff}}}{\sigma_f} - 1 \right)^m \frac{\tau_z}{\sqrt{\tau_r^2 + \tau_\theta^2 + \tau_z^2}} \, H \!\! \left( \frac{\sigma_{\rm{eff}}}{\sigma_f} - 1 \right).
\end{align}
Note that these equations are not all independent since $\lambda_r \lambda_\theta \lambda_z = 1$ implies
\begin{equation}
\frac{\dot{\lambda}_r}{\lambda_r} + \frac{\dot{\lambda}_\theta}{\lambda_\theta}+\frac{\dot{\lambda}_z}{\lambda_z} = 0;
\end{equation}
this is consistent with $\rm{tr}(\bm{\tau})=0$.

\subsubsection{Li diffusion}
For the concentration field, the  conservation equation reads
\begin{equation}
\frac{1}{V_m^B}\frac{\partial c}{\partial t} = -\frac{1}{r}\frac{\partial (rJ_r)}{\partial r}, 
\end{equation}
where $V_m^B$ is the molar volume of Si, $J_r$, the flux of Li, is a function of the chemical potential $\mu$:
\begin{align}
J_r = -\frac{D}{R_gT}\frac{c}{V_m^B}\frac{\partial \mu}{\partial r}, \label{eq:dim_flux}
\end{align}
where $D$ is the diffusivity of Li in Si, $R_g$ is the universal gas constant, and $T$ is the temperature. The chemical potential is further decomposed as
\begin{align}
\mu = \underbrace{\mu_0}_\text{stress-independent} + \underbrace{\mu_S}_\text{stress-dependent}. \label{eq:dim_mu-parts}
\end{align}
The stress-independent part takes the usual form
\begin{equation}
\mu_0 = \mu_0^0 + R_gT\log\left( \gamma c  \right), \label{eq:dim_mu_si}
\end{equation}
where $\mu_0^0$ is a constant representing the chemical potential at a standard state, $\gamma$ is the activity coefficient representing the effects of interactions (leading to non-ideal behaviour) among the atoms/molecules. The stress-dependent part is  taken from \citet{2012JMechPhysSolidsCui} as
\begin{align}
\mu_S &= \frac{V_m^B}{x_{\rm{max}}} \left[ -\frac{1}{3}\frac{\partial J^c}{\partial c} F_{im}^e F_{in}^e C_{mnkl}E_{kl}^e + \frac{1}{2}\left( J^c \frac{\partial C_{ijkl}}{\partial c} + \frac{\partial J^c}{\partial c}C_{ijkl}  \right)E_{ij}^e E_{kl}^e  \right]\label{eq:dim_mu_sd1} \\
&= \frac{V_m^B}{x_{\rm{max}}} \left[ -\frac{1}{3}\frac{\partial J^c}{\partial c} \left\{ \left(\mathbf{F}^e\right)^T \mathbf{F}^e  \right\}:\bm{\sigma}^{0e} + \frac{1}{2} J^c \frac{\partial C_{ijkl}}{\partial c} E_{ij}^e E_{kl}^e + \frac{1}{2} \frac{\partial J^c}{\partial c} \mathbf{E}^e : \bm{\sigma}^{0e} \right] \nonumber \\
& =  \frac{V_m^B}{x_{\rm{max}}} \left[ -\frac{1}{6}\frac{\partial J^c}{\partial c} \mathbf{E}^e :\bm{\sigma}^{0e} - \frac{1}{3}{\rm{tr}}\left( \bm{\sigma}^{0e}  \right)\frac{\partial J^c}{\partial c} + \frac{1}{2} J^c \frac{\partial C_{ijkl}}{\partial c} E_{ij}^e E_{kl}^e \right], \label{eq:dim_mu_sd2}
\end{align}
where we have used $\left(\mathbf{F}^e\right)^T \mathbf{F}^e = 2\mathbf{E}^e + \mathbf{I}$, and where $C$ is the concentration-dependent fourth-rank stiffness tensor. The three terms (within the brackets) of Eq.~\ref{eq:dim_mu_sd2} can be expanded as
\begin{subequations} \label{eq:dim_mu_sd3}
\begin{align}
-\frac{1}{6} \frac{\partial J^c}{\partial c} \mathbf{E}^e : \bm{\sigma}^{0e} &= -\frac{1}{6} \frac{\partial J^c}{\partial c} \frac{E}{(1+\nu)(1-2\nu)} \left[ (1-\nu)\left\{ (E_r^e)^2+ (E_\theta^e)^2 + (E_z^e)^2\right\} \right. \nonumber \\
& \quad \quad \quad \left. + 2\nu \left( E_r^e E_\theta^e + E_\theta^e E_z^e + E_z^e E_r^e \right)  \right], \\
-\frac{1}{3}{\rm{tr}}\left( \bm{\sigma}^{0e} \right)  \frac{\partial J^c}{\partial c} &= -\frac{1}{3} \frac{\partial J^c}{\partial c} \frac{E}{(1+\nu)(1-2\nu)}\left[ (1-\nu)(E_r^e + E_\theta^e + E_z^e) \right. \nonumber \\
& \quad \quad \quad \left. + 2\nu(E_r^e + E_\theta^e + E_z^e )  \right], \\
\frac{1}{2} J^c \frac{\partial C_{ijkl}}{\partial c} E_{ij}^e E_{kl}^e &= \frac{1}{2}J^c \left[ \frac{\partial}{\partial c}\left\{ \frac{E(1-\nu)}{(1+\nu)(1-2\nu)} \right\}  \left\{ (E_r^e)^2 + (E_\theta^e)^2 + (E_z^e)^2 \right\} \right. \nonumber \\
& \quad \quad \quad \left. + 2\frac{\partial}{\partial c} \left\{ \frac{E\nu}{(1+\nu)(1-2\nu)}  \right\} \left( E_r^eE_\theta^e + E_\theta^e E_z^e + E_z^e E_r^e  \right)  \right]. 
\end{align}
\end{subequations}
The activity, $\gamma$, and the diffusivity, D, are assumed to be of the following forms \citep{2012JMechPhysSolidsCui}:
\begin{align}
\gamma = \frac{1}{1-c} \exp \left[ \frac{1}{R_g T} \left\{ 2\left(A_0 - 2B_0\right)c - 3\left( A_0 - B_0  \right)c^2  \right\}  \right], \quad
D = D_0 \exp\left( \frac{\alpha V_m^B \sigma_\theta^0}{R_g T}  \right),
\end{align}
where the values for the parameters $A_0$, $B_0$, $\alpha$, and the concentration-independent coefficient of the diffusivity, $D_0$, are explicitly given in Table~\ref{table:values}.
\subsubsection{Boundary and initial conditions}
We impose free-traction conditions at the radial surface of the cylinder (but not on the top and bottom faces) and no displacement at the centre, that is
\begin{gather}
\sigma_r^0(R_0,t) = 0,\quad u(0,t) = 0.
\end{gather}
For the concentration, we express the boundary condition at the periphery by relating the flux to the linearized version of the Butler-Volmer condition \citep{2012JMechPhysSolidsCui}
\begin{gather}
J_r = \underbrace{J_0 (1-c)}_{\rm{Charging}}, 
\end{gather}
while at the centre of the cylinder, we have
\begin{gather}
J_r = 0.
\end{gather}
On the top and bottom faces of the cylinder, we consider two situations. First, we take the ends to be free; that is, there is no external physical constraint acting on the ends. We model this condition of no physical constraint in the axial direction by requiring the net force acting on the face in the axial direction to be zero. This translates into the integral constraint
\begin{equation}
2\pi  \int_0^{R_0} \sigma_z^0 ~r \; dr = 0. \label{eq:intcon}
\end{equation}
Second, we impose physical constraints that prevent the deformation of the ends of the cylinder in the axial direction. We assume that such physical constraints may be imposed without inhibiting the lateral movement of the ends of the cylinder. That is, we replace Eq.~(\ref{eq:intcon}) by $\partial w/\partial z = 0$ on the faces.

Finally, we take as initial conditions a pristine, stress-free, Li-free electrode, that is
\begin{gather}
\lambda_r(r,0) = \lambda_\theta(r,0) = 1, \quad u(r,0) = 0, \quad c(r,0) = 0.
\end{gather}

\subsection{Non-dimensionalization}
We non-dimensionalize by setting:
\begin{gather}
\tilde{r}=\frac{r}{R_0}, \quad \tilde{z}=\frac{z}{L_0} \quad \tilde{D}=\frac{D}{D_0}, \quad \tilde{t}=\frac{D_0}{R_0^2}t, \quad \tilde{u}=\frac{u}{R_0}, \quad \tilde{w}=\frac{w}{L_0}\nonumber \\ 
\quad \tilde{J}_{r,0} = \frac{V_m^B R_0 }{D_0}J_{r,0}, \quad \tilde{\sigma}_{r,\theta,z,{\rm{eff}},f} = \frac{V_m^B }{R_g T}\sigma_{r,\theta,z,{\rm{eff}},f}, \quad \tilde{\mu}_{0,S} = \frac{1}{R_g T}\mu_{0,S}.
\end{gather}
Here, the undeformed radius, $R_0$, provides a natural length scale for both the radial coordinate, $r$, and the radial displacement field, $u$, while the undeformed length of the cylinder, $L_0$ similarly provides the natural length scale for both the axial coordinate, $z$, and the axial displacement field, $w$. Furthermore, the time required by Li to diffuse a distance equal to the undeformed radius of the cylindrical Si anode, $R_0^2/D_0$, provides for a physically intuitive measure of the time scale. The scale chosen for the influx rate may be justified on the basis of a simple reasoning. An estimate for the number of moles of Li diffused into the Si may be made using the concentration measure of Li, $c$, and the molar volume of Si, $V_m^B$; therefore, the concentration in terms of number of moles for a unit length of the cylinder scales as $R_0^2/V_m^B$. Since the influx rate is measured per unit surface area, per unit time, for the flux scale we need the surface area measure, $R_0$ (again per unit length of the cylinder), and the time scale which we have already chosen as $R_0^2/D_0$ -- thus, giving us a scale for the flux as $(R_0^2/V_m^B)(1/R_0)(D_0/R_0^2)$, that is $D_0/(V_m^B R_0)$.
In terms of this non-dimensionalization scheme, the governing equations become:
\begin{align}
\frac{\partial c}{\partial \tilde{t}} & = -\frac{\partial \tilde{J}_r}{\partial \tilde{r}} - \frac{\tilde{J}_r}{\tilde{r}}, \label{eq:nondim_c}\\ 
0 &= \frac{\partial \tilde{\sigma}_r^0}{\partial \tilde{r}} + \frac{\tilde{\sigma}_r^0 - \tilde{\sigma}_\theta^0}{\tilde{r}}, \label{eq:nondim_sigma} \\
\frac{\partial \lambda_r}{\partial \tilde{t}} &= \sqrt{\frac{3}{2}}\lambda_r \dot{d}_0 \frac{R_0^2}{D_0} \left( \frac{\tilde{\sigma}_{\rm{eff}}}{\tilde{\sigma}_f} - 1\right)^m \frac{\tilde{\tau}_r}{ \sqrt{\tilde{\tau}_r^2 + \tilde{\tau}_\theta^2 + \tilde{\tau}_z^2} } \; H\left( \frac{\tilde{\sigma}_{\rm{eff}}}{\tilde{\sigma}_f} - 1 \right), \label{eq:nondim_lambdar}\\
\frac{\partial \lambda_\theta}{\partial \tilde{t}} &= \sqrt{\frac{3}{2}}\lambda_\theta \dot{d}_0 \frac{R_0^2}{D_0} \left( \frac{\tilde{\sigma}_{\rm{eff}}}{\tilde{\sigma}_f} - 1\right)^m \frac{\tilde{\tau}_\theta}{ \sqrt{\tilde{\tau}_r^2 + \tilde{\tau}_\theta^2 + \tilde{\tau}_z^2} } \; H\left( \frac{\tilde{\sigma}_{\rm{eff}}}{\tilde{\sigma}_f} - 1 \right) \label{eq:nondim_lambdath}.
\end{align}
In Eq.~(\ref{eq:nondim_sigma}), the Piola-Kirchhoff stresses may be expressed non-dimensionally in terms of the displacement field by first relating them to the strains using Eq.~(\ref{eq:dim_PK-strains}) as:
\begin{subequations} \label{eq:nondim_PK-strains}
\begin{align}
\tilde{\sigma}_r^0 &= J^c \frac{\tilde{E}(c)}{(1+\nu)(1-2\nu)} \left[ (1-\nu)E_r^e + \nu \left( E_\theta^e + E_z^e  \right)  \right] \frac{2E_r^e + 1}{1+\partial \tilde{u}/\partial \tilde{r}}, \\
\tilde{\sigma}_\theta^0 &= J^c \frac{\tilde{E}(c)}{(1+\nu)(1-2\nu)} \left[ (1-\nu)E_\theta^e + \nu \left( E_z^e + E_r^e  \right)  \right] \frac{2E_\theta^e + 1}{1+ \tilde{u}/\tilde{r}}, \\
\tilde{\sigma}_z^0 &= J^c \frac{\tilde{E}(c)}{(1+\nu)(1-2\nu)} \left[ (1-\nu)E_z^e + \nu \left( E_r^e + E_\theta^e  \right)  \right] \frac{2E_z^e + 1}{1+\partial \tilde{w}/\partial \tilde{z}},
\end{align}
\end{subequations}
where $\tilde{E}(c) = V_m^B E_0 (1+\eta_E x_{\rm{max}}c)/(R_g T)$ is the non-dimensional modulus of elasticity, and then relating the strains to the displacement using Eq.~(\ref{eq:dim_strain-displacement}) as
\begin{subequations} \label{eq:nondim_strain-displacement}
\begin{align}
E_r^e &= \frac{1}{2} \left[ \left( F_r^e \right)^2 - 1 \right] = \frac{1}{2} \left(J^c\right)^{-2/3} \frac{\left( 1+\partial \tilde{u}/\partial \tilde{r} \right)^2}{\lambda_r^2} - \frac{1}{2}, \\
E_\theta^e &= \frac{1}{2} \left[ \left( F_\theta^e \right)^2 - 1 \right] = \frac{1}{2} \left(J^c\right)^{-2/3} \frac{\left( 1+\tilde{u}/\tilde{r} \right)^2}{\lambda_\theta^2} - \frac{1}{2}, \\
E_z^e &= \frac{1}{2} \left[ \left( F_z^e \right)^2 - 1 \right] = \frac{1}{2} \left(J^c\right)^{-2/3} \frac{\left( 1+\partial \tilde{w}/\partial \tilde{z} \right)^2}{\lambda_z^2} - \frac{1}{2}.
\end{align}
\end{subequations}
In Eq.~(\ref{eq:nondim_c}) the non-dimensional flux is given by
\begin{align} 
\tilde{J}_r = - \tilde{D} c \frac{\partial \tilde{\mu}}{\partial \tilde{r}}, \label{eq:nondim_flux}
\end{align}
where
%
%
%
%
%
%
%
\begin{gather}
\tilde{\mu} = \frac{\mu_0^0}{R_g T} + \log(\gamma c) + \tilde{\mu}_{S1} + \tilde{\mu}_{S2} + \tilde{\mu}_{S3},
\end{gather}
with
\begin{subequations}
\begin{align}
\tilde{\mu}_{S1} &=  -\frac{1}{6 x_{\rm{max}}} \frac{\partial J^c}{\partial c} \frac{\tilde{E}}{(1+\nu)(1-2\nu)} \left[ (1-\nu) \left\{ (E_r^e)^2 + (E_\theta^e)^2 + (E_z^e)^2 \right\} \right. \nonumber \\
& \left. \qquad +2\nu (E_r^e E_\theta^e + E_\theta^e E_z^e + E_z^e E_r^e)  \right], \\
\tilde{\mu}_{S2} &= -\frac{1}{3 x_{\rm{max}}} \frac{\partial J^c}{\partial c} \frac{\tilde{E}}{(1+\nu)(1-2\nu)} \left[ (1+\nu) (E_r^e + E_\theta^e + E_z^e) \right], \\ 
\tilde{\mu}_{S3} &= \frac{1}{2 x_{\rm{max}}} J^c \left[ \frac{\partial}{\partial c} \left\{ \frac{\tilde{E}(1-\nu)}{(1+\nu)(1-2\nu)} \right\} \left\{ (E_r^e)^2 + (E_\theta^e)^2 + (E_z^e)^2  \right\} \right. \nonumber \\ 
& \qquad \left. + 2 \frac{\partial}{\partial c} \left\{ \frac{\tilde{E}\nu}{(1+\nu)(1-2\nu)}  \right\} (E_r^eE_\theta^e + E_\theta^eE_z^e + E_z^eE_r^e)  \right].
\end{align}
\end{subequations}
Eqs.~(\ref{eq:nondim_lambdar}) and (\ref{eq:nondim_lambdath}) are expressed in terms of the non-dimensionalized deviatoric parts of the Cauchy stress tensor given by
\begin{align}
\tilde{\tau}_{r,\theta,z} = \tilde{\sigma}_{r,\theta,z} - \frac{1}{3}(\tilde{\sigma}_r + \tilde{\sigma}_\theta + \tilde{\sigma}_z),
\end{align}
where
\begin{subequations} \label{eq:nondim_Cauchy}
\begin{align}
\tilde{\sigma}_r &= \frac{\tilde{E}}{(1+\nu)(1-2\nu)}\left[ (1-\nu)E_r^e + \nu\left( E_\theta^e + E_z^e  \right)  \right] \frac{\sqrt{2E_r^e + 1}}{\sqrt{2E_\theta^e + 1}\sqrt{2E_z^e + 1}} \\
\tilde{\sigma}_\theta &= \frac{\tilde{E}}{(1+\nu)(1-2\nu)}\left[ (1-\nu)E_\theta^e + \nu\left( E_z^e + E_r^e  \right)  \right] \frac{\sqrt{2E_\theta^e + 1}}{\sqrt{2E_z^e + 1}\sqrt{2E_r^e + 1}} \\
\tilde{\sigma}_z &= \frac{\tilde{E}}{(1+\nu)(1-2\nu)}\left[ (1-\nu)E_z^e + \nu\left( E_r^e + E_\theta^e  \right)  \right] \frac{\sqrt{2E_z^e + 1}}{\sqrt{2E_r^e + 1}\sqrt{2E_\theta^e + 1}}.
\end{align}
\end{subequations}
Furthermore, the non-dimensional effective stress in Eqs.~(\ref{eq:nondim_lambdar}) and (\ref{eq:nondim_lambdath}) is given following Eq.~(\ref{eq:dim_effectivestress}) by
\begin{align}
\tilde{\sigma}_{\rm{eff}} = \sqrt{\frac{3}{2}} \sqrt{\tilde{\tau}_r^2 + \tilde{\tau}_\theta^2 + \tilde{\tau}_z^2}.  \label{eq:nondim_effectivestress}
\end{align}
The boundary conditions and the initial condition corresponding to Eq.~(\ref{eq:nondim_c}) are:
\begin{gather}
c(\tilde{r},0)=0,\quad \tilde{J}_r(0,\tilde{t}) = 0, \quad \tilde{J}_r(1,\tilde{t})=\underbrace{\tilde{J}_0(1-c)}_{\rm{Charging}} \;\; \text{or} \;\; \underbrace{-\tilde{J}_0 c}_{\rm{Discharging}}, 
\end{gather}
The boundary conditions corresponding to Eq.~(\ref{eq:nondim_sigma}) are:
\begin{gather}
\tilde{u}(0,\tilde{t}) = 0, \quad \tilde{\sigma}_r^0(1,\tilde{t}) = 0.
\end{gather}
The initial conditions corresponding to Eqs.~(\ref{eq:nondim_lambdar}) and (\ref{eq:nondim_lambdath}) are:
\begin{gather}
\lambda_r(\tilde{r},0)=1, \quad \text{and} \quad \lambda_\theta(\tilde{r},0)=1.
\end{gather}
When the ends of the cylinder are constrained we have no axial strain, $\partial w /\partial z = 0$. When the ends are free we have instead no axial force,
\begin{gather}
2 \pi \int_0^{1} \tilde{\sigma}_z^0  \tilde{r} \; d\tilde{r}=0.
\end{gather}

\subsection{Buckling criteria} \label{subsec:special}

In axially constrained case, the natural propensity of the cylinder to expand in the axial direction during lithiation, generates a compressive axial force. As this compressive force increases, it ultimately leads to a critical situation in which the cylinder buckles. To study this, we choose the simplest possible buckling criterion based on the classical Euler column buckling formula, which was shown to be valid asymptotically for thin enough cylinders \citep{dedego11},
\begin{equation}
F_{\rm{crit}} = \frac{\pi^2 EI}{(KL)^2},\label{eq:Fcr}
\end{equation}
where $F_{\rm{crit}}$ is the critical net force in the axial direction that induces buckling. Here $L$ is the length of the column, $EI$ is the flexural rigidity (where $E$ is the modulus of elasticity and $I$ is the second moment of area), and $K$ is a factor which accounts for the conditions at the boundary. Assuming the cylinder is constrained between two frictionless parallel plates -- which most closely represents an actual battery condition for the cylindrical electrode particle -- we have $K=1$. The effect of growth on the Euler buckling criterion is to induce both a compressive residual stress and a change of geometry. Here, we will only consider the dominant contribution of the geometry as the residual stresses remain bounded by the yield stress. There may also be a modification of the criterion due to the plastic deformation but such a derivation is beyond the scope of this work and we expect that Eq.~(\ref{eq:Fcr}) represents a reasonable first approximation.  

Notwithstanding this simplifying assumption, we account for two phenomena associated with the swelling cylinder but not considered in the classical buckling theory. First, the radius used for the calculation of the second moment of area is changing, and so we have, in dimensional terms,
\begin{gather}
{R} = {R}_0 \left( 1 + \frac{\partial {u}}{\partial {r}} \right).
\end{gather}
This alone brings about a significant modification to the classical Euler buckling criterion because the second moment of area, which depends on the radius through the relation
\begin{gather}
I = \frac{\pi R^4}{4},
\end{gather}
now changes with time. We will refer to this as the first modification of the Euler buckling criterion.

Second, we need to account for the variation in the modulus of elasticity due to its dependence on the Li concentration, and, hence, the variations of the flexural rigidity with position through the cross-section since the concentration itself varies spatially. To incorporate this consistently, however, is not trivial. We recognize that the appearance of the flexural rigidity in Eq.~(\ref{eq:Fcr}) is due fundamentally to an integral over the bending stresses distributed throughout the column cross-section. In our case, the bending stress is simply the stress, $\sigma_z$, in the axial direction. Following the classical route to account for the moment due to the bending stress, we first note that $\sigma_z = -\kappa E r$, where $\kappa$ is the local curvature of the column; see \citet[Chap. 4]{HowellBook}. We then calculate the moment as
\begin{gather}
M = \int_{-R}^{R} \breve{r} \sigma_z dA,
\end{gather}
where $\breve{r}$ is the radial coordinate in the current frame of reference, the elemental area, $dA = 2\sqrt{R^2-\breve{r}^2}d\breve{r}$, and the integration is carried out over the entire current circular cross-section of the cylindrical anode. From this, the time-dependent flexural rigidity with concentration-dependent, spatially varying modulus of elasticity may be extracted as:
\begin{align}
EI(c,t) &=4 \int_{0}^{R} \breve{r}^2  E \sqrt{R^2-\breve{r}^2} \; d\breve{r} \\
	&= 4 R^4 \int_{0}^{1} \tilde{r}^2 E_0 \left( 1 + \eta_E x_{max} c \right) \sqrt{1-\tilde{r}^2} \; d\tilde{r}.
\end{align}
We will refer to this as the second modification of the Euler buckling criterion. 

\begin{table}[t!] 
\caption{} \label{table:values}
\centering
\begin{tabular}{l l}
\hline \hline
Material property or parameter & Value \\
\hline
$A_0$, parameter used in activity constant & -29549 Jmol$^{-1\;a}$ \\
$B_0$, parameter used in activity constant & -38618 Jmol$^{-1\;a}$ \\
$D_0$, diffusivity of Si & 1 $\times$ 10$^{-16}$ m$^2$s$^{-1\;b}$ \\
$\dot{d}_0$, characteristic strain rate for plastic flow in Si & 1 $\times 10^{-3}$ s$^{-1\;a}$ \\
$E_0$, modulus of elasticity of pure Si & 90.13 GPa$^{\;c}$ \\
$m$, stress exponent for plastic flow in Si & 4$^{\;d}$ \\
$R_g$, universal gas constant & 8.314 JK$^{-1}$mol$^{-1}$ \\
$R_0$, initial radius of unlithiated Si electrode & 200 nm \\
$T$, temperature & 300 K \\
$V_m^B$, molar volume of Si & 1.2052 $\times$ 10$^{-5}$ m$^3$mol$^{-1\;a}$ \\
$x_{\rm{max}}$, maximum concentration of Li in Si & 4.4 \\
$\alpha$, coefficient of diffusivity & 0.18$^{\;e}$ \\
$\eta$, coefficient of compositional expansion & 0.2356$^{\;a}$ \\
$\eta_E$, rate of change of modulus of elasticity with concentration & -0.1464$^{\;c}$ \\
$\nu$, Poisson's ratio of Si & 0.28$^{\;a}$ \\
$\sigma_f$, initial yield stress of Si & 0.12 GPa$^{\;d}$ \\
\hline \\
$^a$ \citet{2012JMechPhysSolidsCui} \\
$^b$ \citet{2011NanoLettLiu} \\
$^c$ \citet{2010JElectrochemSocRhodes} \\
$^d$ \citet{2011JMechPhysSolidsBower} \\
$^e$ \citet{2011JPowerSourcesHaftbaradaran}
\end{tabular}
\end{table}

\section{Results and Discussions} \label{sec:resdisc}
We  solve numerically the system of equations as described in Sec.~\ref{sec:formulation} using the finite element package COMSOL Multiphysics. For the solution we use the values of the material properties and the parameters as shown in Table~\ref{table:values}. We first present the results corresponding to the axially unconstrained case. We then present the axially constrained case, discussing in detail the critical behaviour associated with buckling. In the presentation of the results, we use the term ``capacity" to mean specific capacity with units of mAhg$^{-1}$. This is calculated as: Capacity $= \left( \int_0^{R_0} c \,2\pi r \; dr \right) / \left( \int_0^{R_0} 2 \pi r \; dr \right) $, and is a measure of the state of charge of the electrode with respect to its reference configuration. 

\subsection{Axially unconstrained} \label{subsec:generalresults}
\begin{figure}
\centering
\begin{subfigure}[b]{0.45\textwidth}
\includegraphics[width=\textwidth]{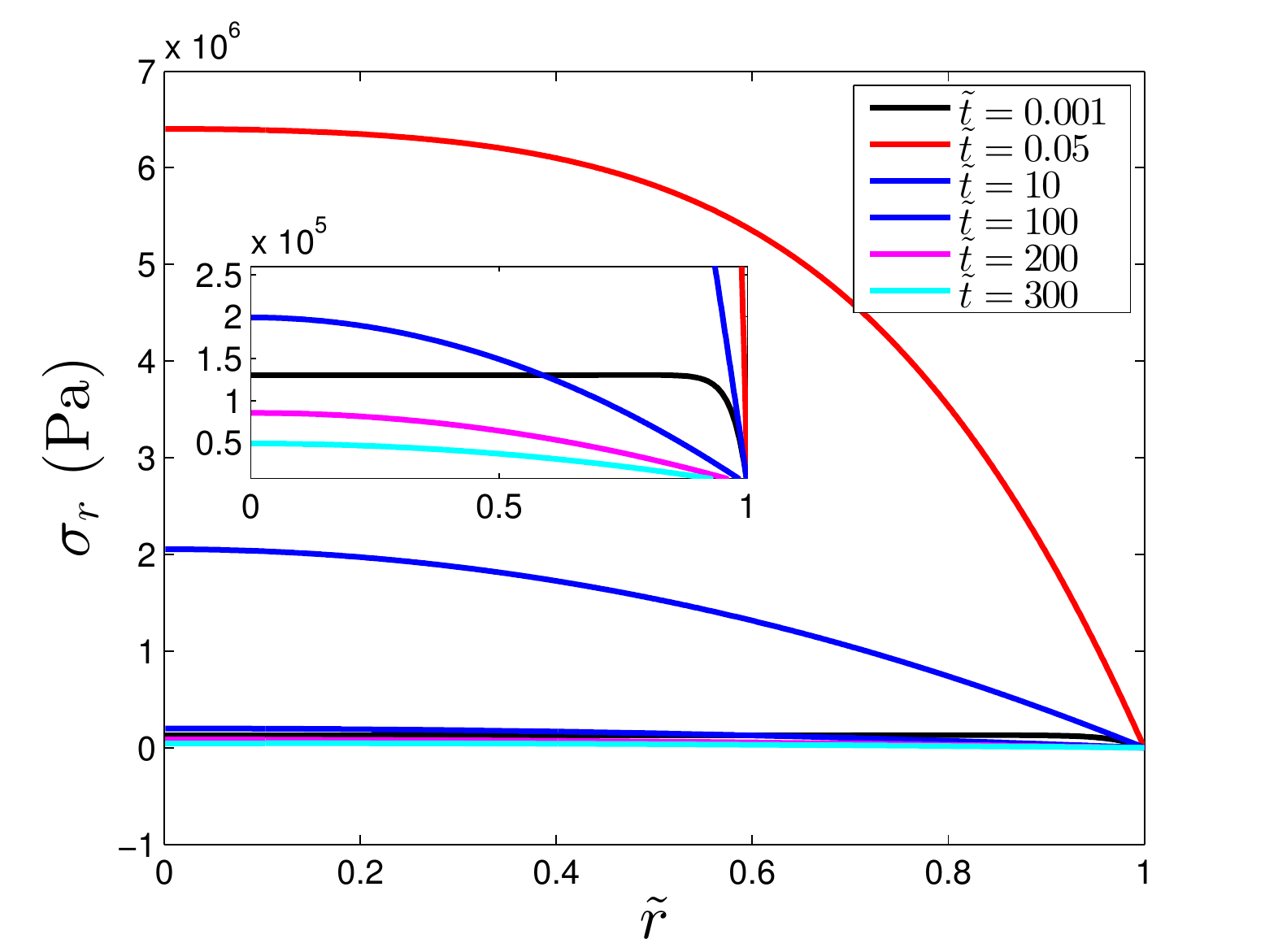}
\caption{}
\end{subfigure}%
\begin{subfigure}[b]{0.45\textwidth}
\includegraphics[width=\textwidth]{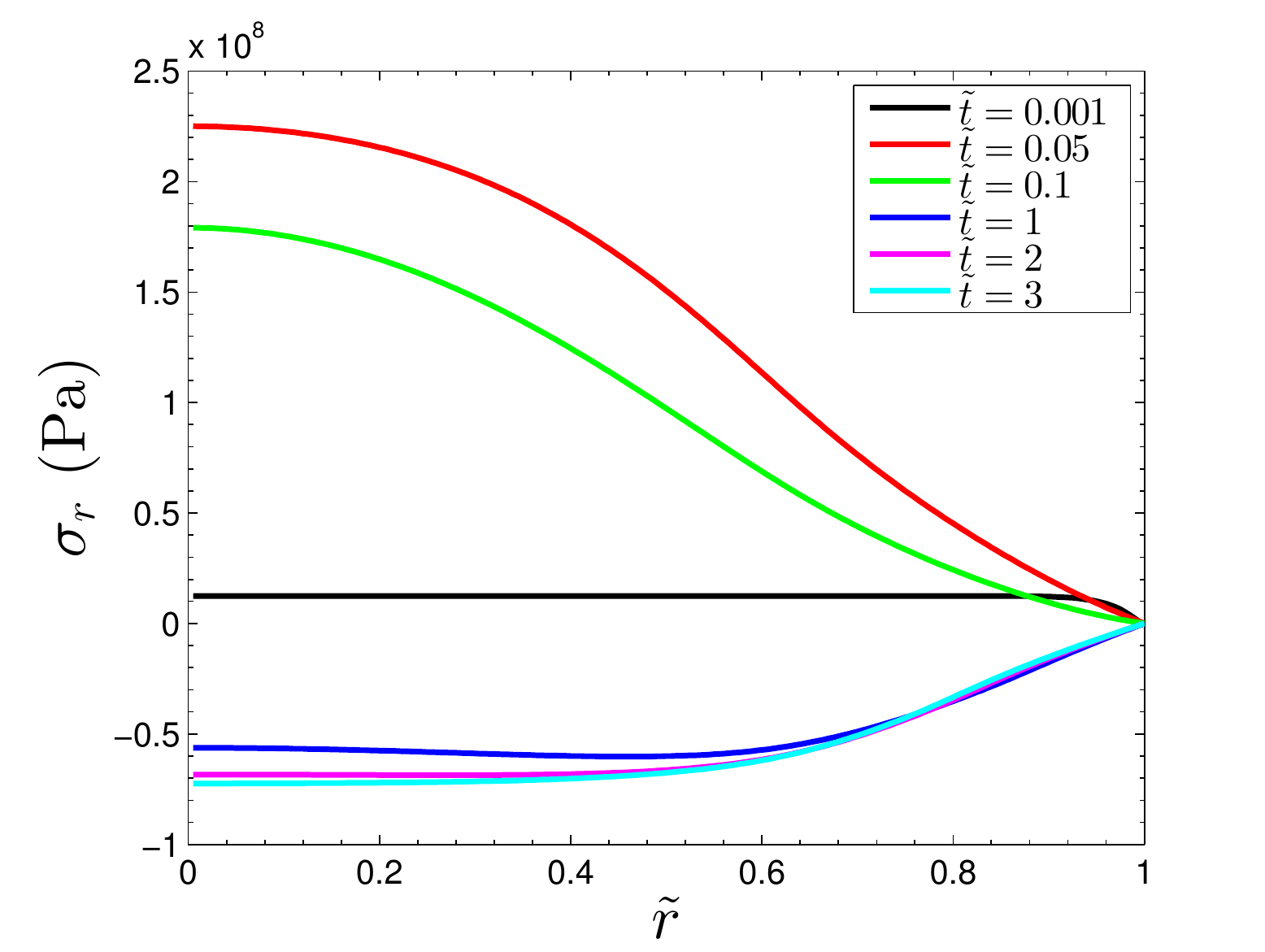}
\caption{}
\end{subfigure}
\begin{subfigure}[b]{0.45\textwidth}
\includegraphics[width=\textwidth]{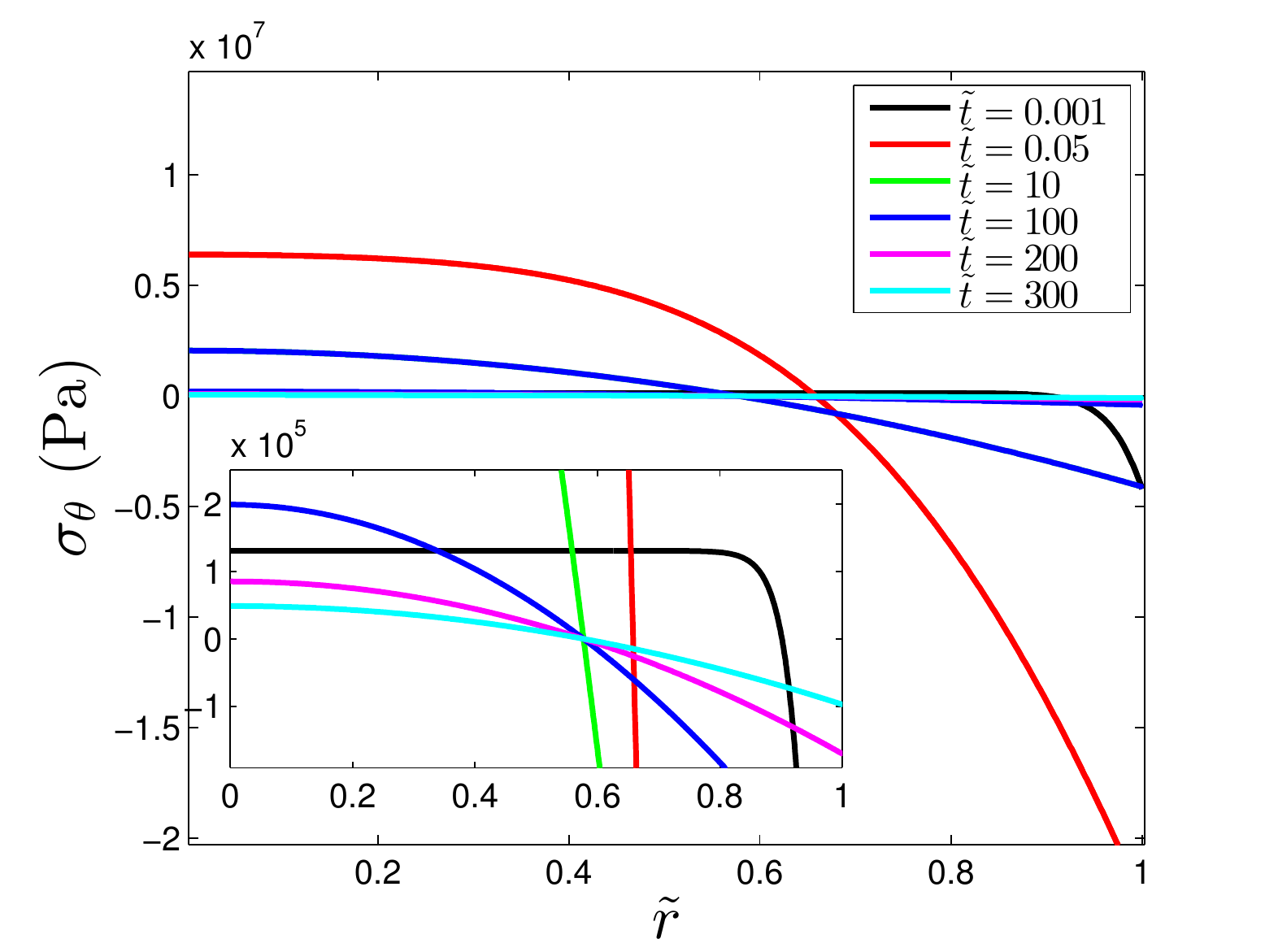}
\caption{}
\end{subfigure}%
\begin{subfigure}[b]{0.45\textwidth}
\includegraphics[width=\textwidth]{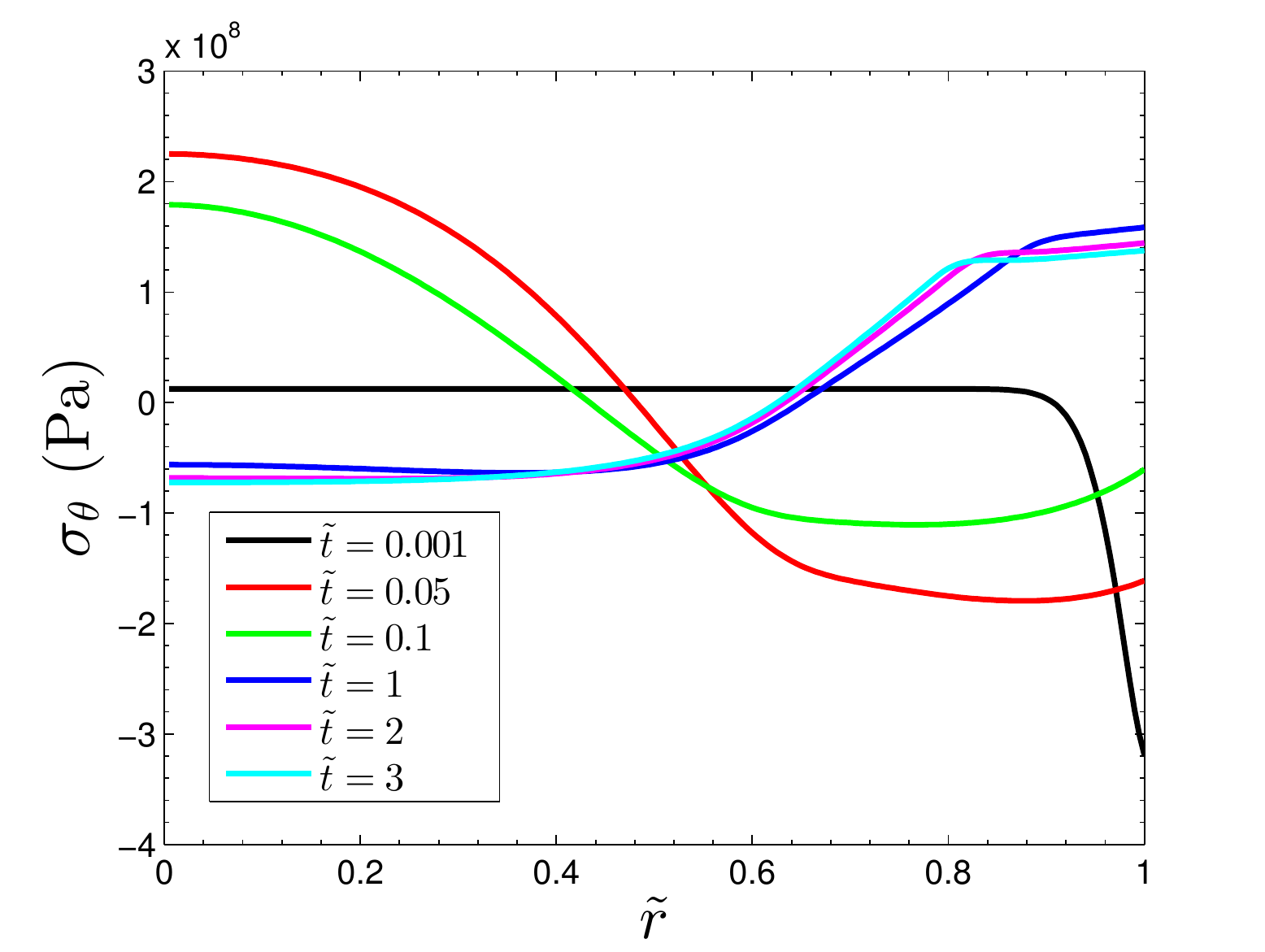}
\caption{}
\end{subfigure}
\begin{subfigure}[b]{0.45\textwidth}
\includegraphics[width=\textwidth]{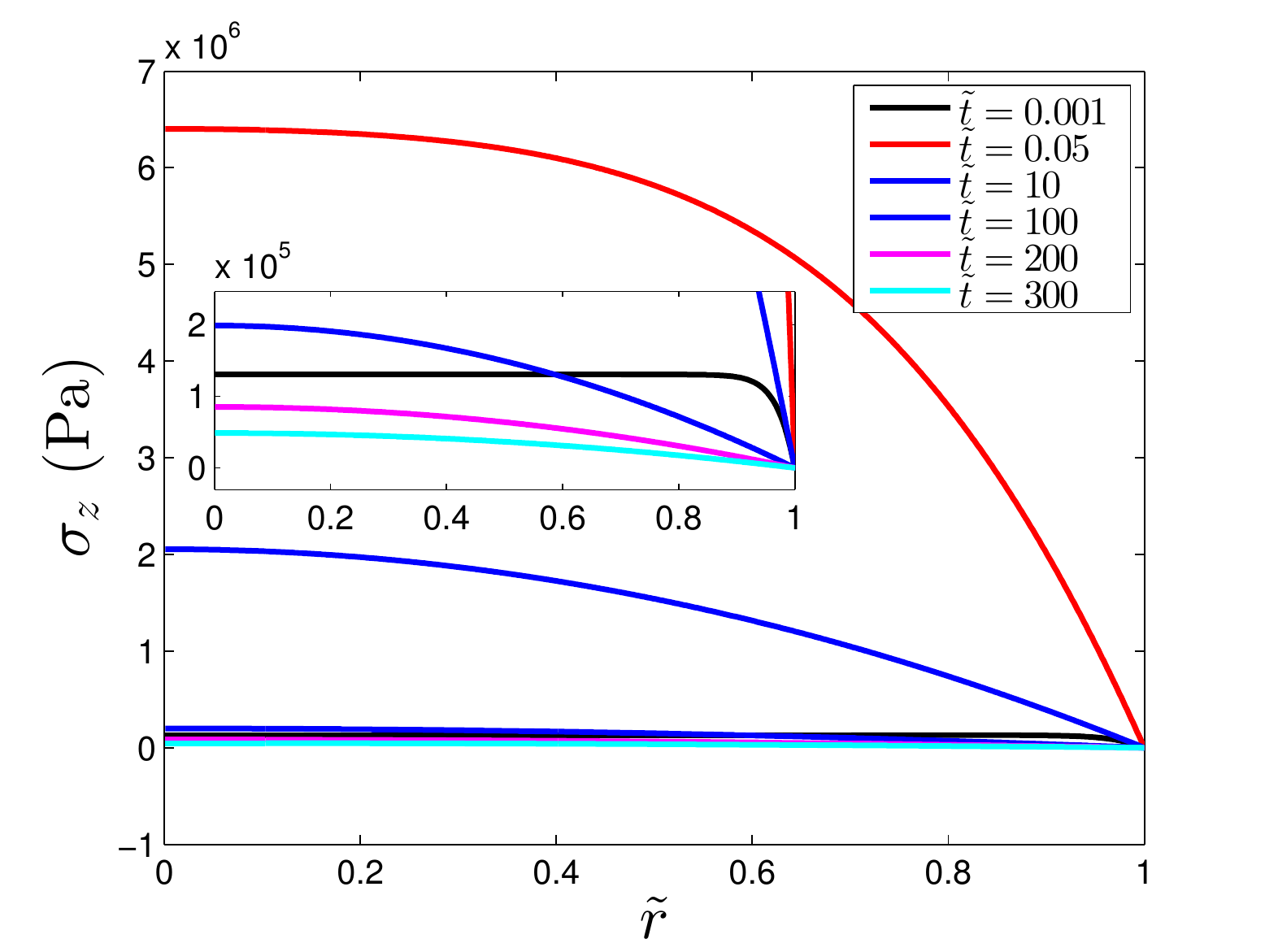}
\caption{}
\end{subfigure}%
\begin{subfigure}[b]{0.45\textwidth}
\includegraphics[width=\textwidth]{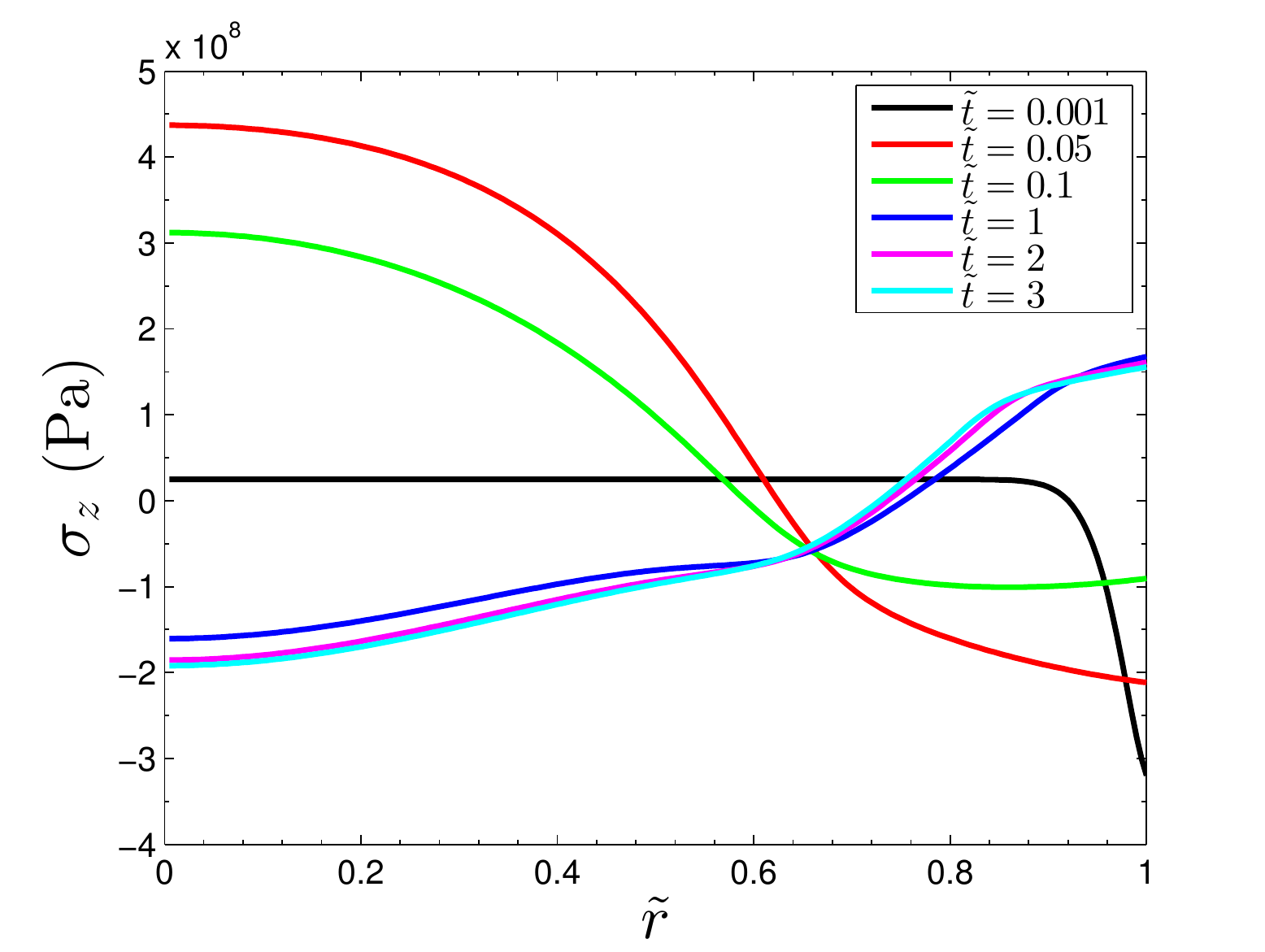}
\caption{}
\end{subfigure} 
\caption{Evolution of stress (represented dimensionally in units of Pa) with time for the case when no physical constraint is imposed in the axial direction. Panels (a), (c), and (e) correspond to a charging condition with (a) $\tilde{J}_0=0.001$, while (b), (d), and (f) to that with $\tilde{J}_0=0.1$. Insets in (a), (c), and (e) show zoomed-in views at higher times.} \label{fig:intcon_sigma}
\end{figure}
\begin{figure}[h!]
\centering
\begin{subfigure}[b]{0.45\textwidth}
\includegraphics[width=\textwidth]{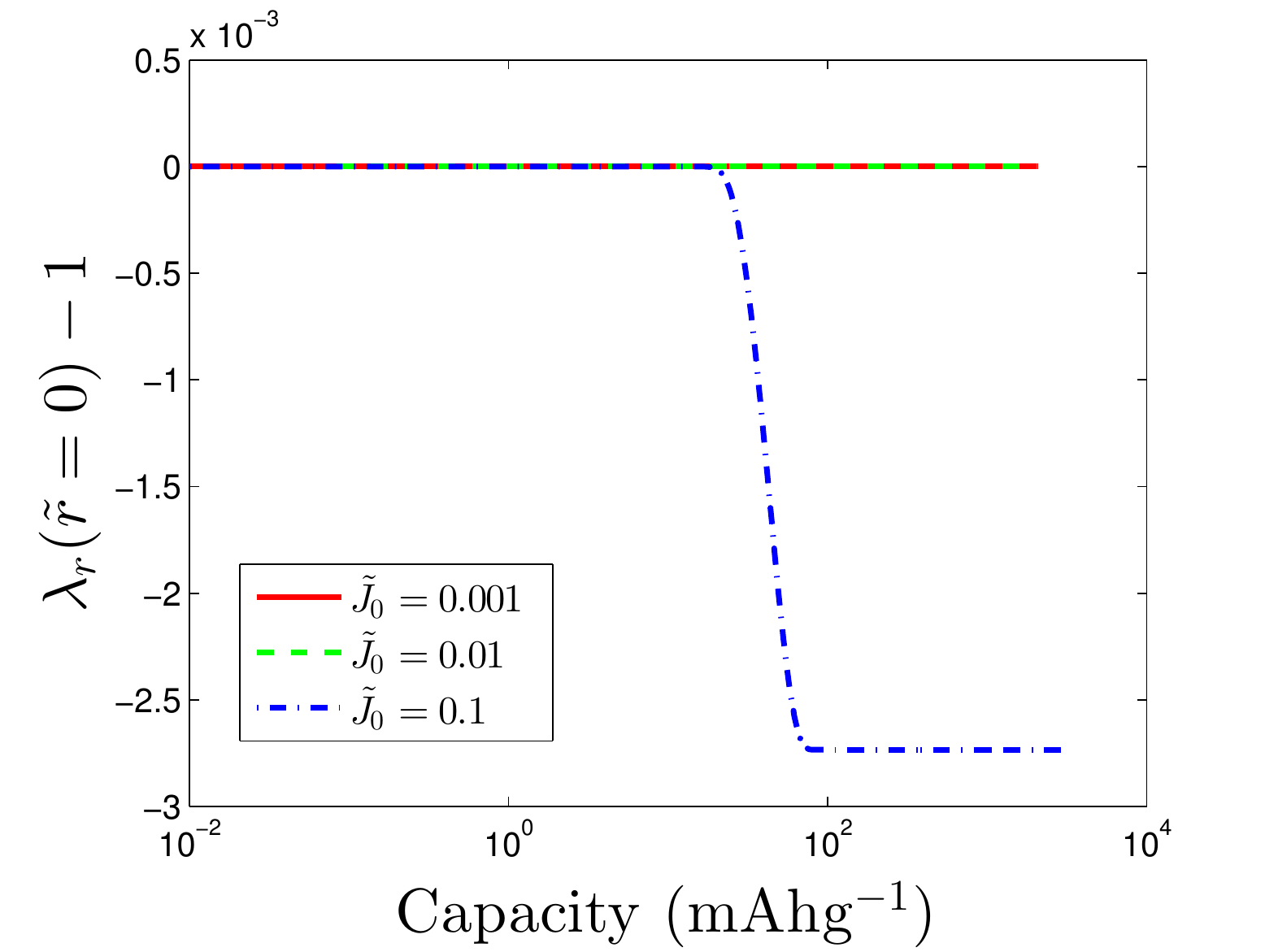}
\caption{}
\end{subfigure}%
\begin{subfigure}[b]{0.45\textwidth}
\includegraphics[width=\textwidth]{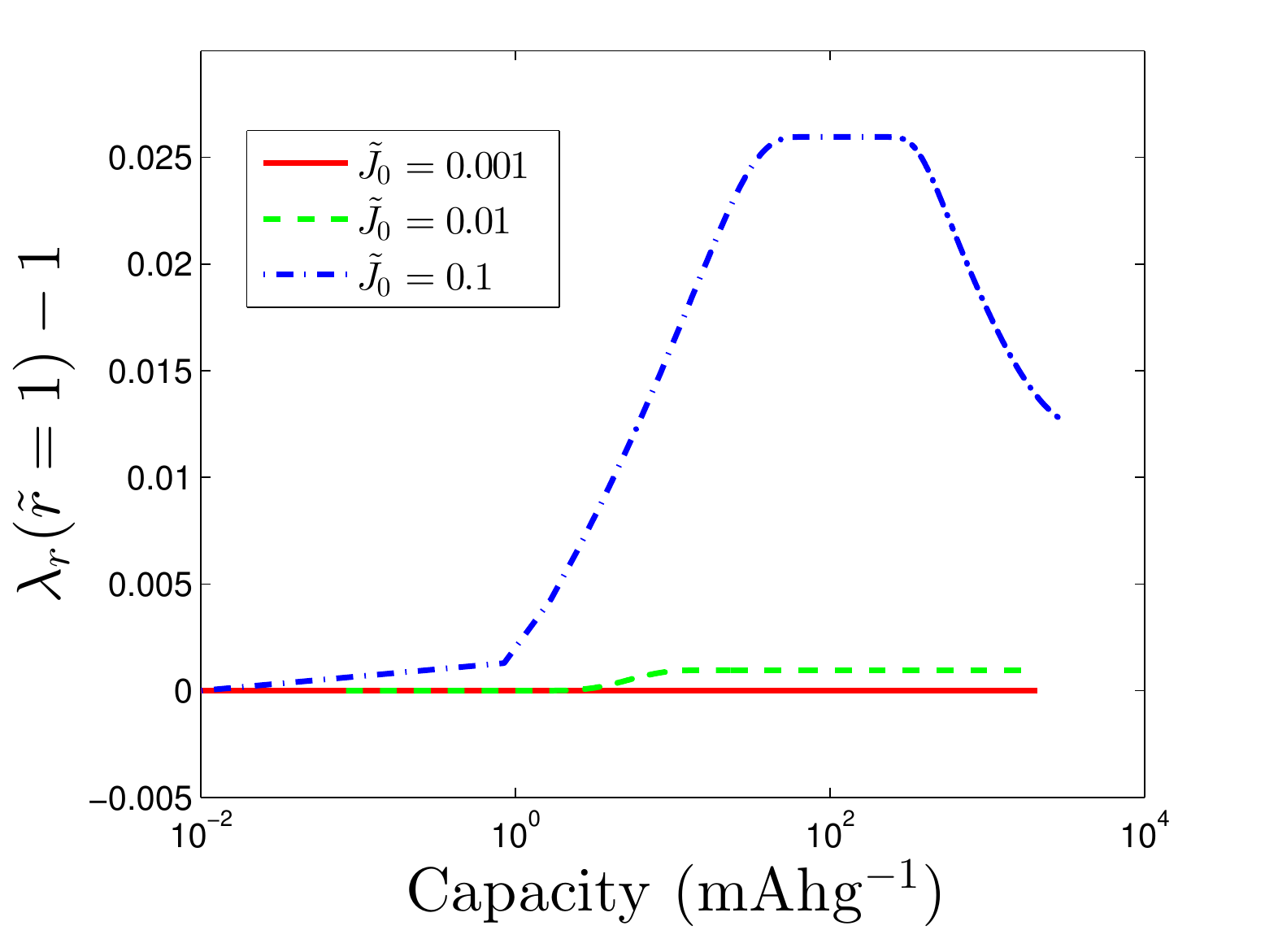}
\caption{}
\end{subfigure}
\begin{subfigure}[b]{0.45\textwidth}
\includegraphics[width=\textwidth]{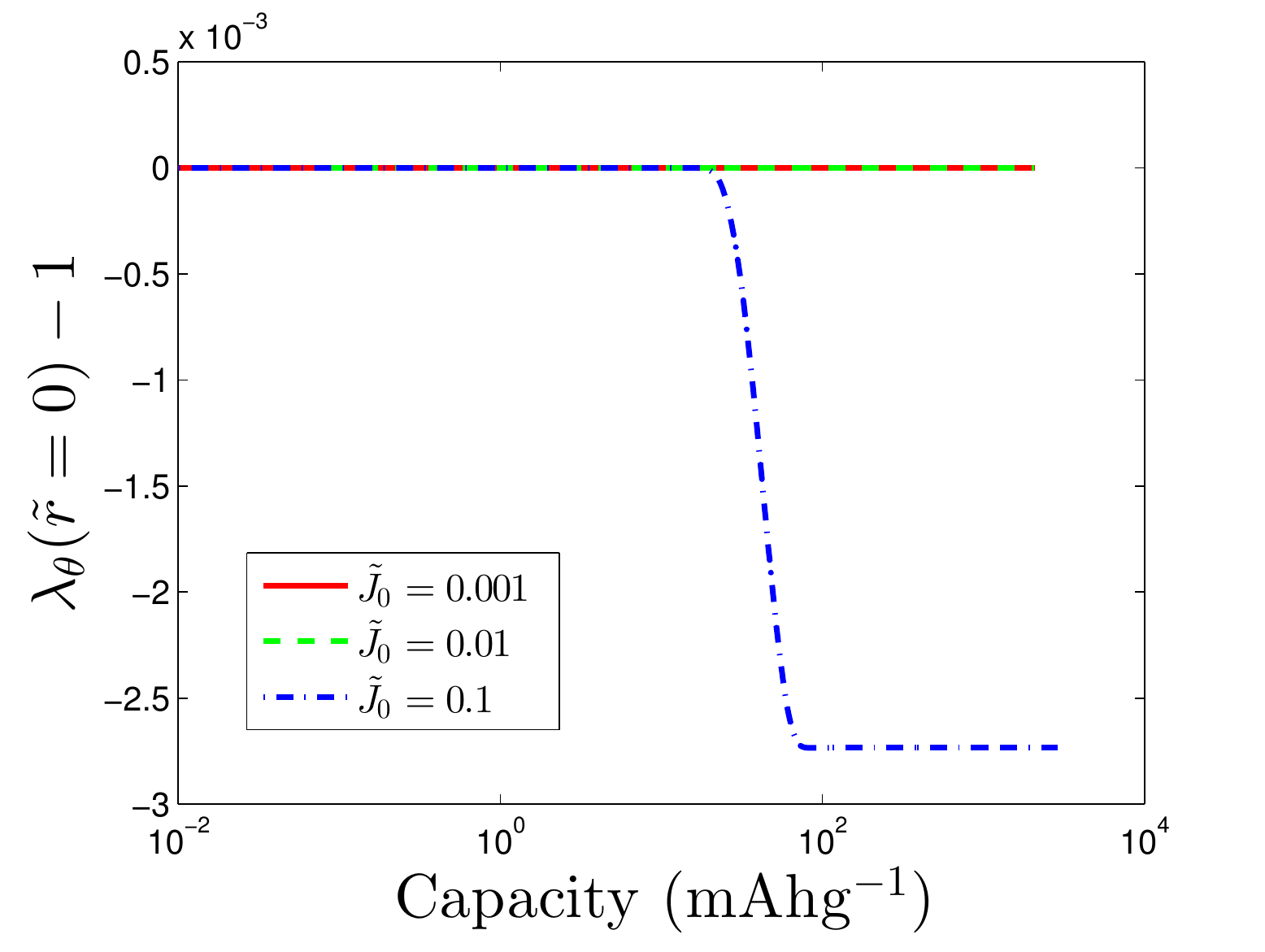}
\caption{}
\end{subfigure}%
\begin{subfigure}[b]{0.45\textwidth}
\includegraphics[width=\textwidth]{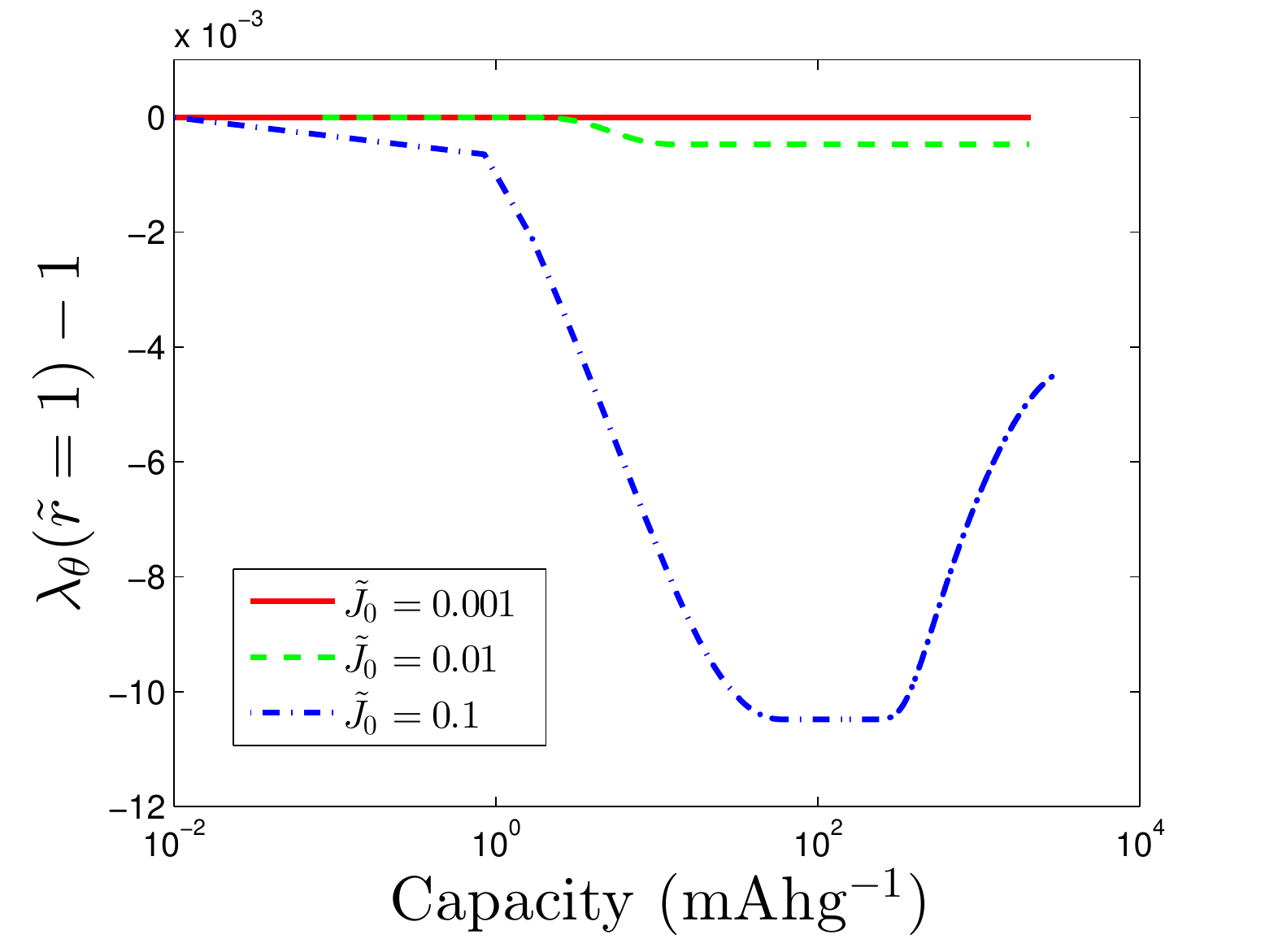}
\caption{}
\end{subfigure}
\begin{subfigure}[b]{0.45\textwidth}
\includegraphics[width=\textwidth]{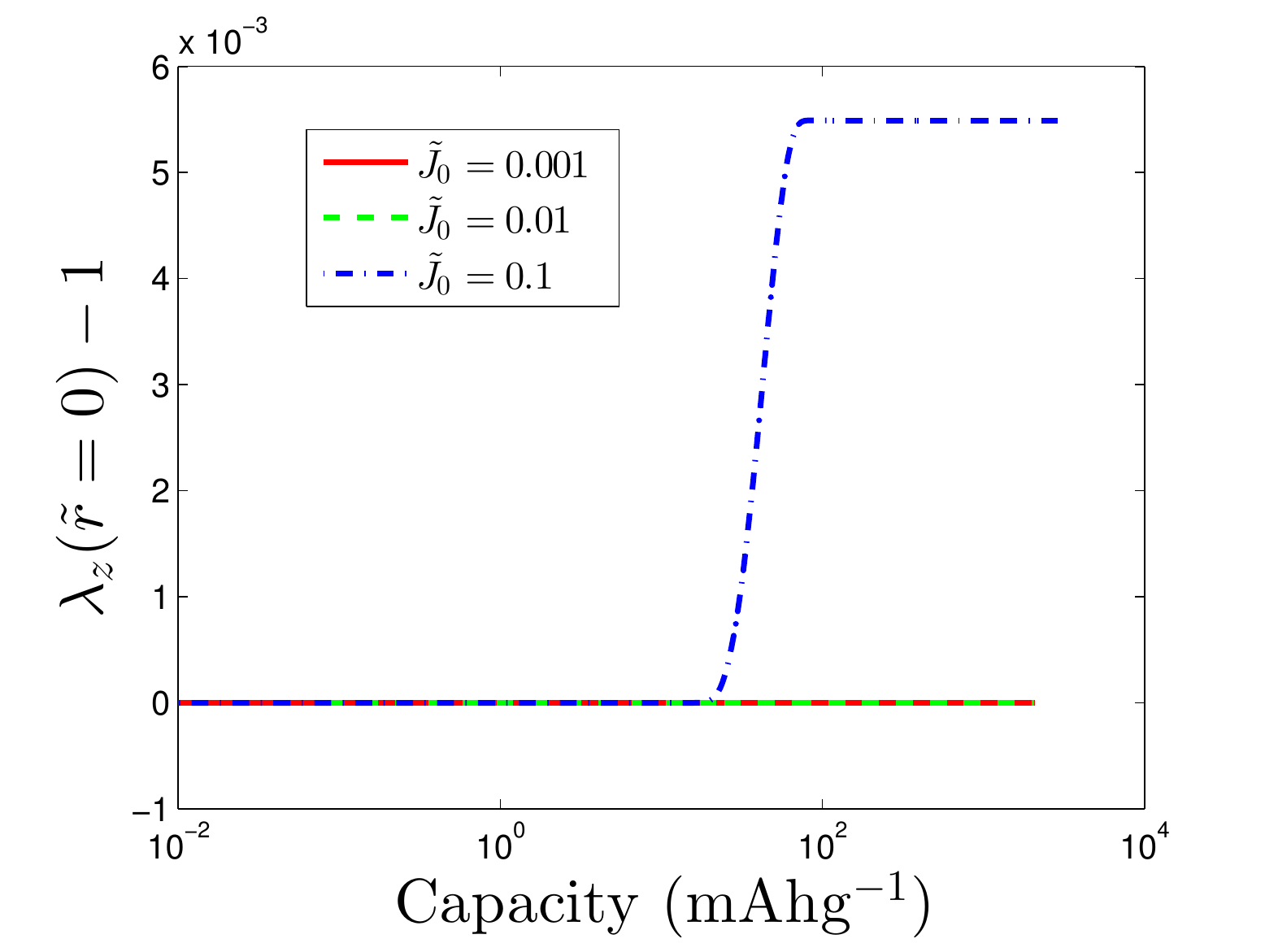}
\caption{}
\end{subfigure}%
\begin{subfigure}[b]{0.45\textwidth}
\includegraphics[width=\textwidth]{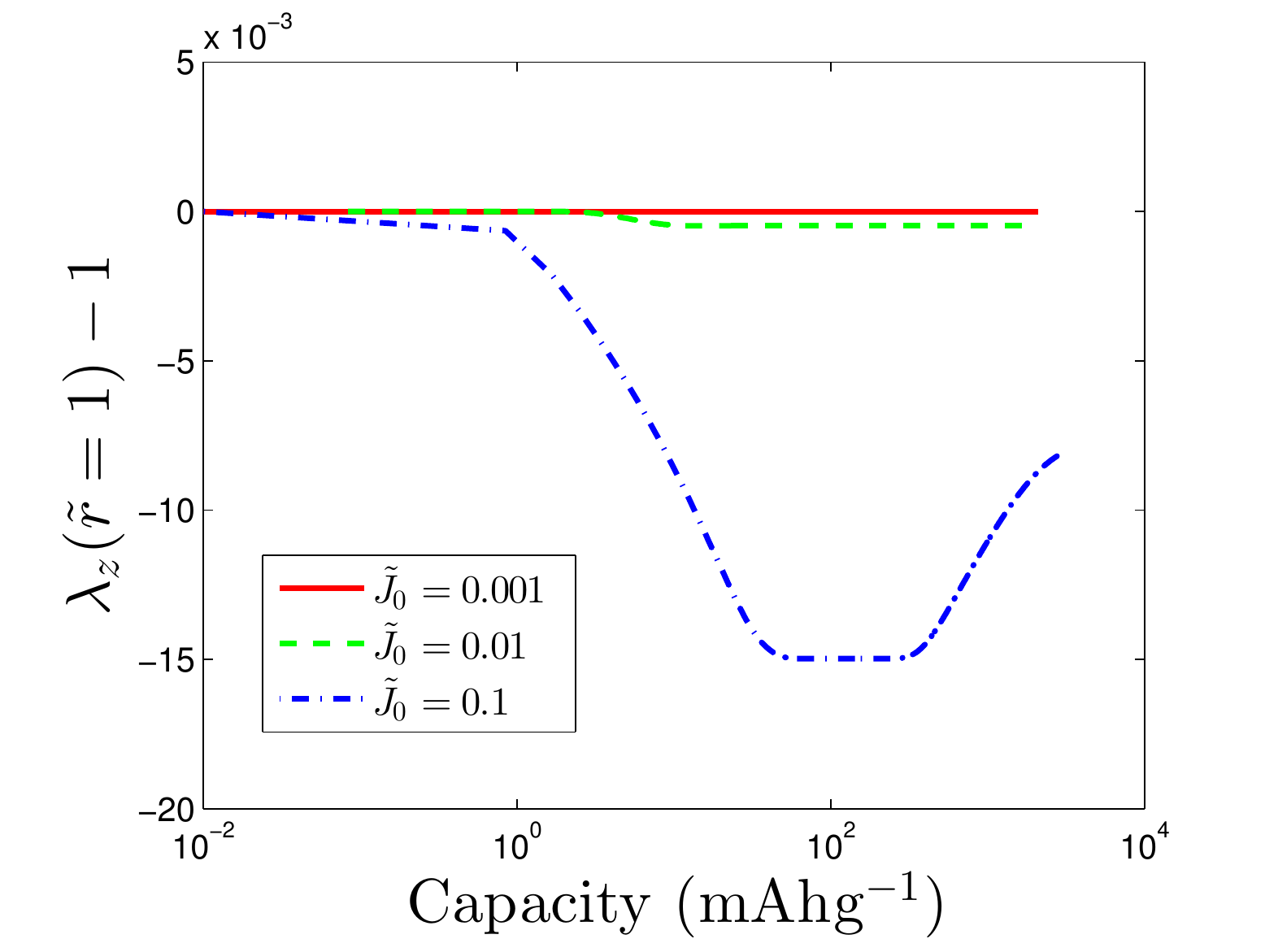}
\caption{}
\end{subfigure} 
\caption{Variation of plastic stretches with increasing lithium concentration represented in terms of the capacity (expressed dimensionally in units of mAhg$^{-1}$) for the case when no physical constraint is imposed in the axial direction. Panels (a), (c), and (e) correspond to the centre ($\tilde{r}=0$), while (b), (d), and (f) to the surface ($\tilde{r}=1$).} \label{fig:intcon_lambda}
\end{figure}

Fig.~\ref{fig:intcon_sigma} shows the variation of the stress along the radial, hoop, and axial directions over the radius of the unconstrained cylinder corresponding to two different lithiation rates, $\tilde{J}_0=0.001$ and $\tilde{J}_0=0.1$. The most interesting trends are observed for $\tilde{J}_0=0.1$. We now provide a  physical interpretation of the observed trends. As lithiation begins at the relatively high influx rate of $\tilde{J}_0=0.1$, it is only the silicon near the surface which becomes lithiated. With increasing lithium concentration, there is a tendency for volumetric growth. Not being subjected to any constraint radially from the outside (the traction-free boundary condition), tensile stresses are induced in the radial direction just beneath the surface, and these tensile stresses are propagated all the way to the centre. In the hoop and the axial directions, however, the presence of adjacent material points acts as a self-constraint, and thus the tendency for volumetric growth induces compressive stresses in those directions for some distance beneath the surface. However, closer to the centre the radial and the hoop stresses  become indistinguishable. Therefore,  closer to the centre, the hoop stresses are also tensile in nature. Note that the physical boundary conditions in the axial direction are not the same as the hoop direction and the condition we have used is that of zero net force which translates into an integral constraint over the stresses.

With time, lithium diffuses further inside the cylinder, and this changes the nature of the stresses significantly. The volumetric growth further induces compressive stresses in the hoop direction near the surface due to the self-constraint imposed by the presence of adjacent material points. Therefore,  compressive stresses are  induced along the three principal  directions for points close to the axis. Growth inside the cylinder naturally pushes central point against the material points closer to the surface which are already in a state of compressive stress as discussed previously. This gradually relaxes the compressive stresses near the surface, and ultimately reverts the stress state there to a tensile one both along the hoop and the axial direction. 

The evolution of the stresses for the lower influx rate, $\tilde{J}_0$, is significantly slower than that for $\tilde{J}_0=0.1$ as clearly seen in Fig.~\ref{fig:intcon_sigma} (a), (c), and (e). The non-dimensional time, $\tilde{t}=300$ translates into a dimensional time of more than 33 hours for the choice of the parameter values given in Table~\ref{table:values}. It is seen from the insets of Fig.~\ref{fig:intcon_sigma} (a), (c), and (e) that even after this high charging time, the stresses do not revert in nature vis-\`{a}-vis what is observed in Fig.~\ref{fig:intcon_sigma} (b), (d), and (f) for the higher influx rate. An important point to note here is that the `state of charge' in terms of the specific capacity is practically the same ($\approx 1800$ mAhg$^{-1}$) for the $\tilde{J}_0=0.001$ case at $\tilde{t}=300$ as that for the $\tilde{J}_0=0.1$ one at $\tilde{t} = 3$.

Fig.~\ref{fig:intcon_lambda} shows the variation of plastic stretches at the centre (panels (a), (c), and (e)) and at the surface (panels (b), (d), and (f)) as lithium concentration builds up inside the anode; this concentration is represented in terms of the capacity. An important observation is that the plastic stretches at the centre deviate from 1, indicating that plastic deformation does indeed occur \emph{throughout} the entire cylindrical cross-section. This is in sharp contrast to the spherical case investigated by \citet{2012JMechPhysSolidsCui} where it was found that even for a relatively high charging rate corresponding to $\tilde{J}_0=0.1$ there was no plastic deformation at the centre of the sphere throughout the charging time; in other words, the elastic-plastic front did not propagate into the centre. This contrast may be  understood from the yield criterion used. In this regard, we note that at the centre, $\sigma_r = \sigma_\theta$, and it is true in both the present cylindrical case as well as in the spherical case. However, the additional symmetry leading to the equivalence of the stresses in the polar and the azimuthal directions due to the very geometry of the spherical case ensures that an almost hydrostatic state of stress is present at the sphere centre - ensuring, thus, that the effective stress (this is what is actually used in the yield criterion) value is always zero there. In the present cylindrical case, however, this additional symmetry is absent. Thus on the axis, $\sigma_z \neq \sigma_r = \sigma_\theta $. Additionally, we note that the effective stress from Eq.~(\ref{eq:dim_effectivestress}) may be rewritten as 
\begin{equation}
\sigma_{\rm{eff}} = \sqrt{\frac{1}{2}} \sqrt{\left(\sigma_r-\sigma_\theta\right)^2 + \left(\sigma_\theta - \sigma_z\right)^2 + \left(\sigma_z-\sigma_r\right)^2},
\end{equation}
from which it is clear that $\sigma_{\rm{eff}}$ does not vanish at the centre.

 Fig.~\ref{fig:intcon_sigma} illustrates other trends in the evolution of the plastic stretches at the centre and at the surface. First, at the start of lithiation, tensile plastic stretches in the radial direction are induced at the surface in keeping with the discussion related to the radial stresses. Similarly, in the hoop and in the axial directions, compressive plastic stretches are also induced during these initial times. As lithium diffuses inside the cylinder with time, the development of compressive stresses in the radial and the hoop directions at the centre leads to compressive plastic stretches along both those directions. This necessarily requires the plastic stretches in the axial direction at the centre to be tensile as plastic deformations are isochoric. At this relatively advanced stage of lithiation, there is a significant change in the nature of the plastic stretches at the surface. It is observed that along  the three principal directions, the evolution of the plastic stretches, tensile in the radial direction, and compressive in the hoop and the axial directions, plateau off, and then reverse direction altogether. This is consistent with the previous physical picture of the material points in the inner regions of the cylinder pushing out against the surface due to growth with increasing lithium concentration, and in the process, reverting the stress-states near the surface. 

\subsection{Axially constrained} \label{subsec:specialresults}
\begin{figure}
\centering
\begin{subfigure}[b]{0.45\textwidth}
\includegraphics[width=\textwidth]{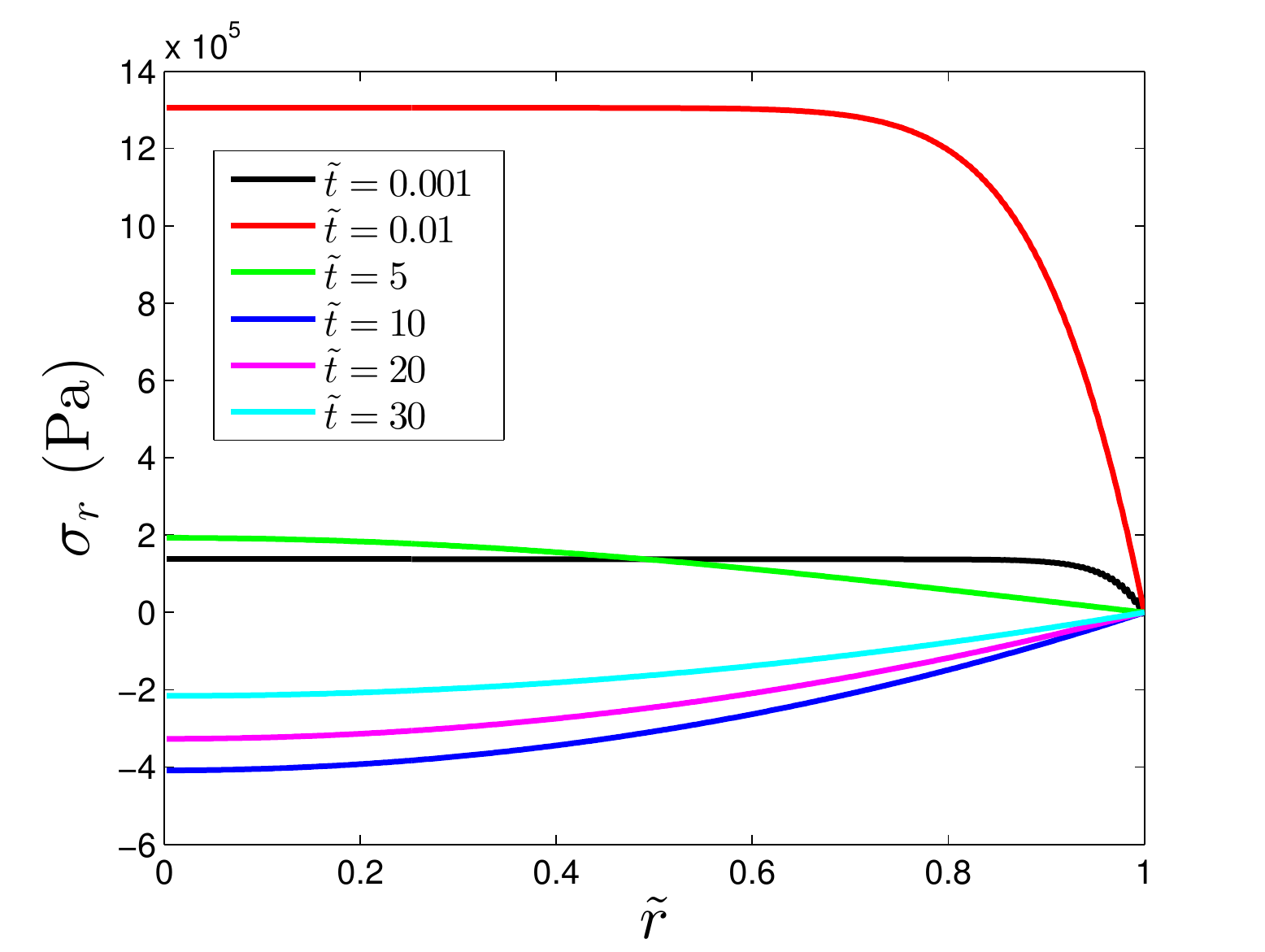}
\caption{}
\end{subfigure}%
\begin{subfigure}[b]{0.45\textwidth}
\includegraphics[width=\textwidth]{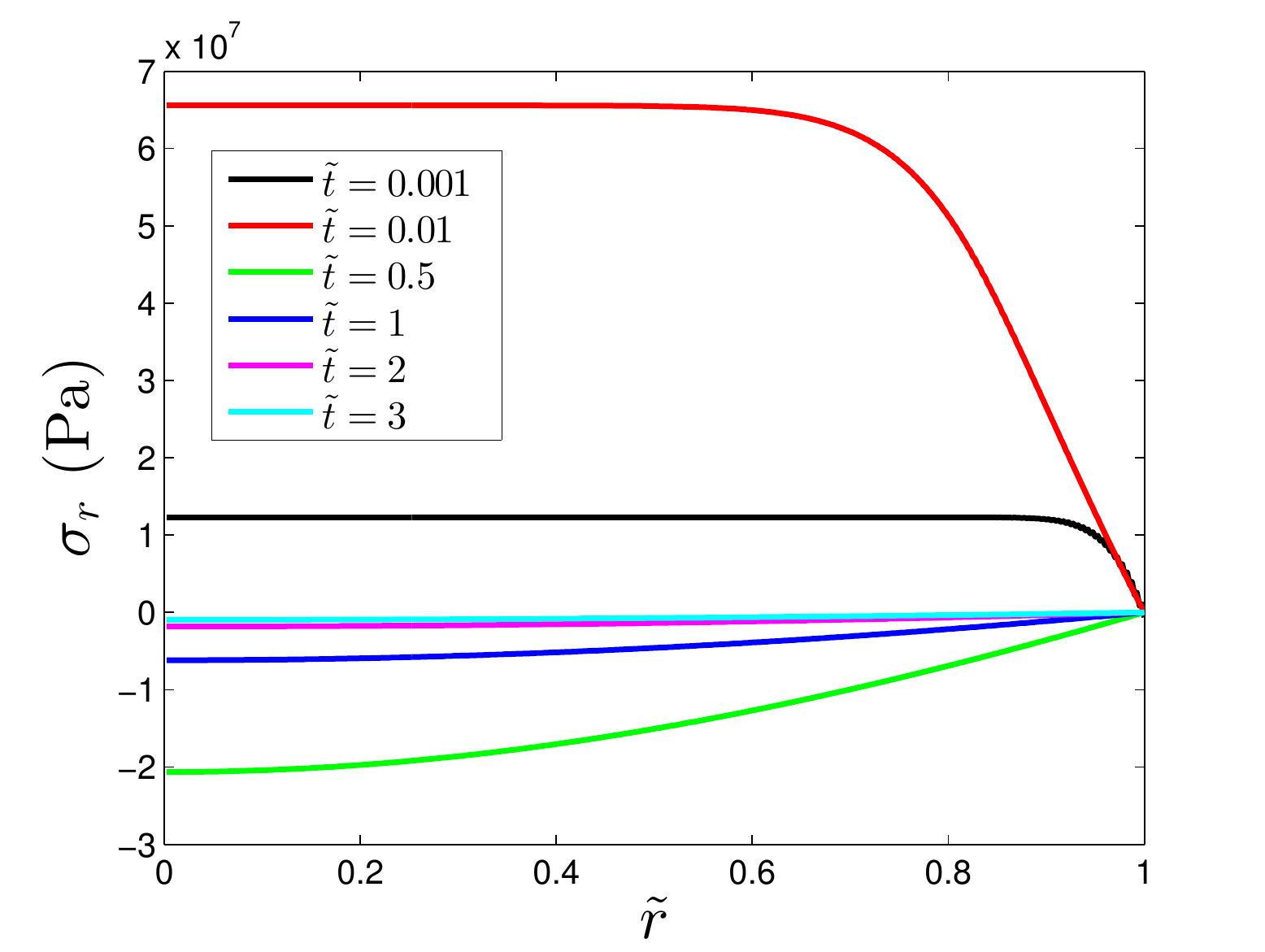}

\caption{}
\end{subfigure}
\begin{subfigure}[b]{0.45\textwidth}
\includegraphics[width=\textwidth]{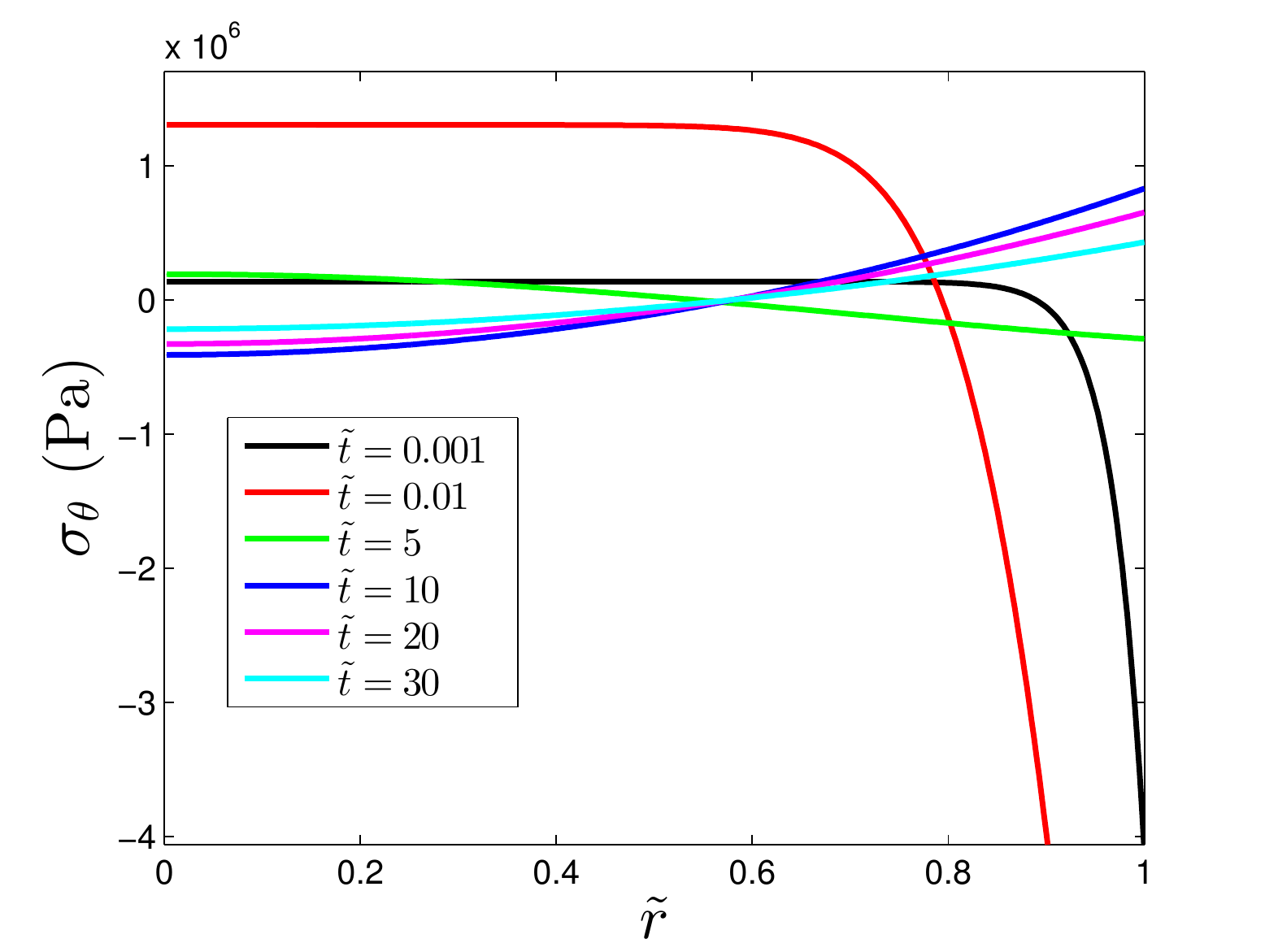}
\caption{}
\end{subfigure}%
\begin{subfigure}[b]{0.45\textwidth}
\includegraphics[width=\textwidth]{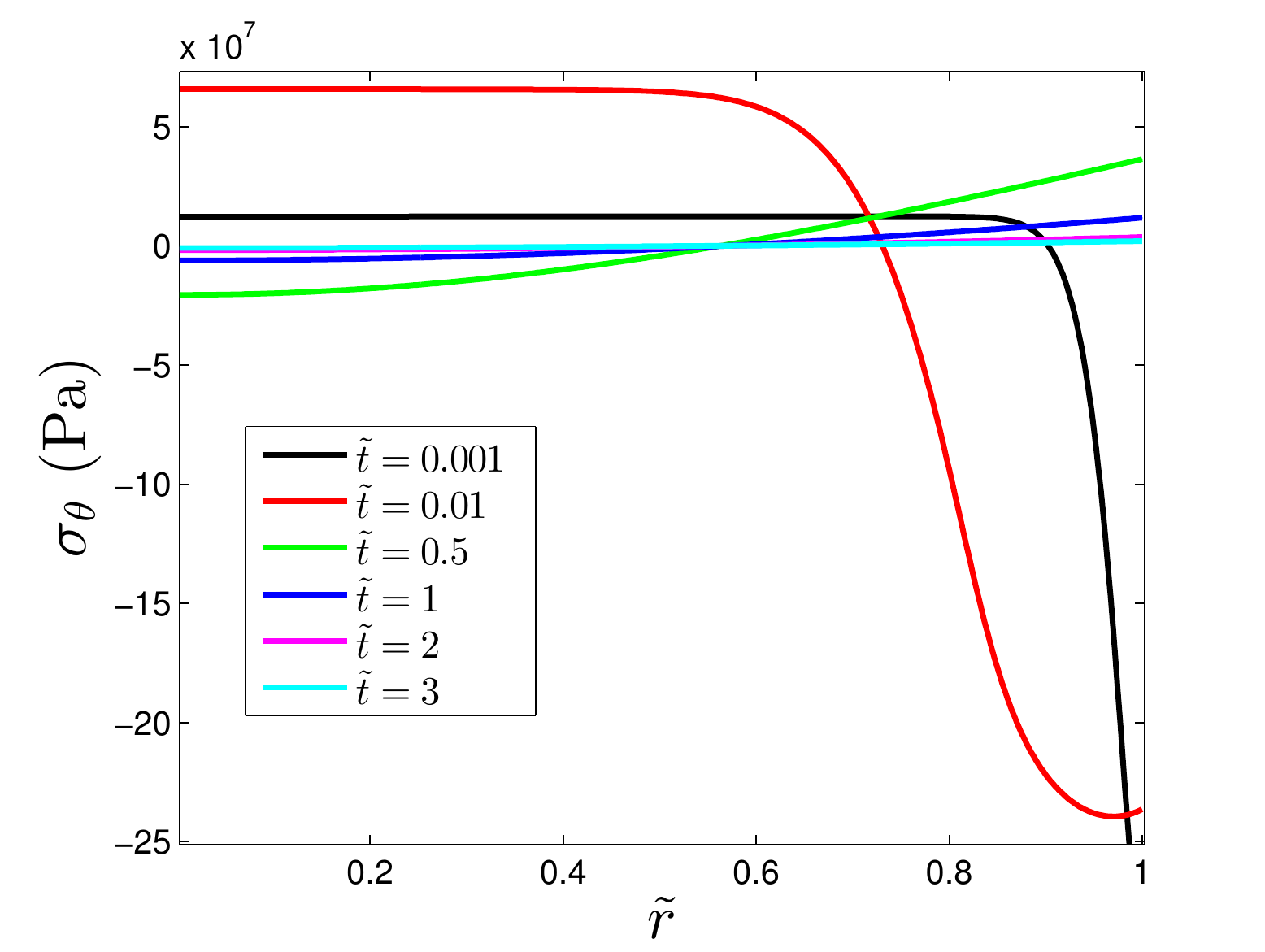}
\caption{}
\end{subfigure}
\begin{subfigure}[b]{0.45\textwidth}
\includegraphics[width=\textwidth]{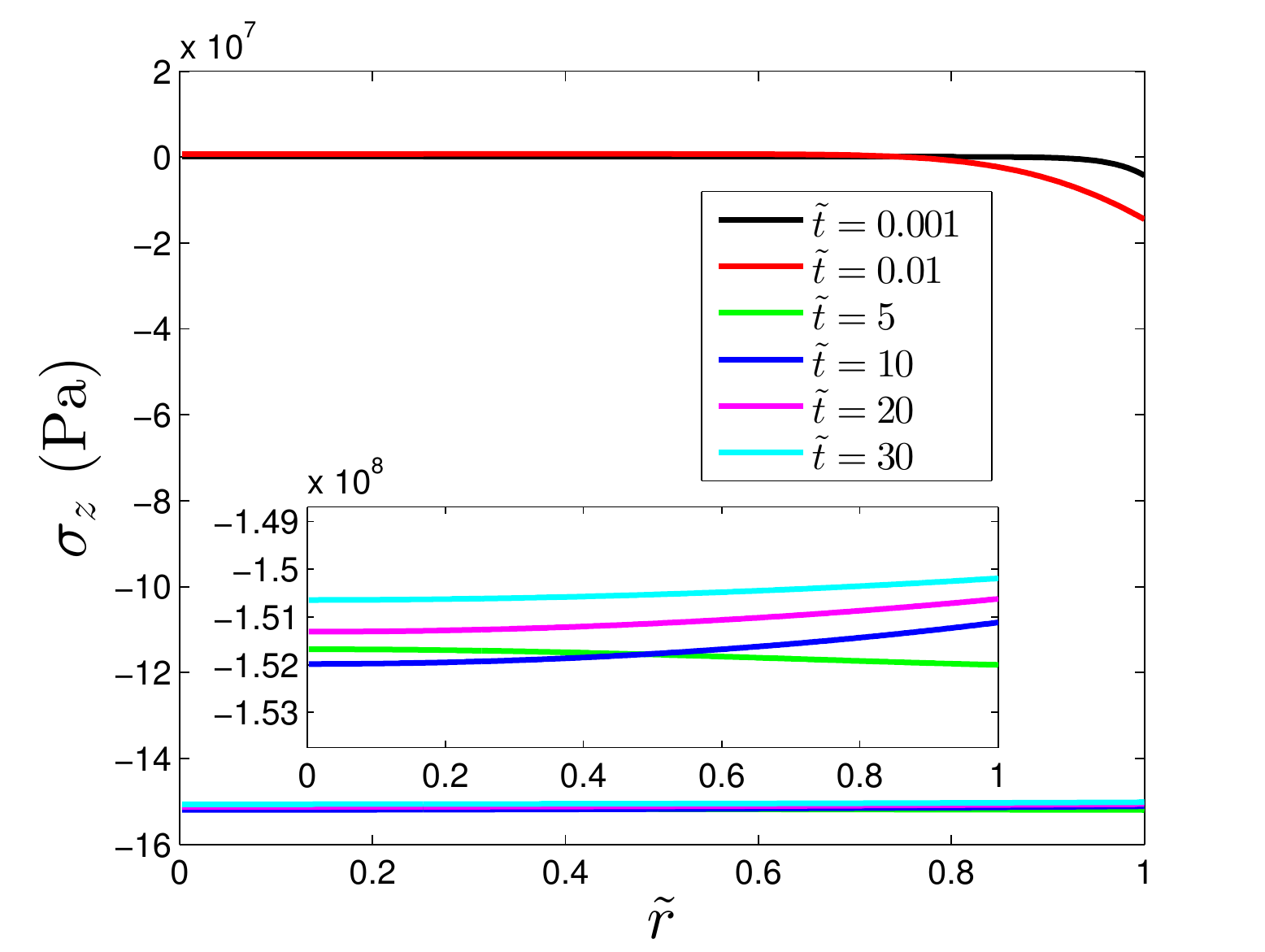}
\caption{}
\end{subfigure}%
\begin{subfigure}[b]{0.45\textwidth}
\includegraphics[width=\textwidth]{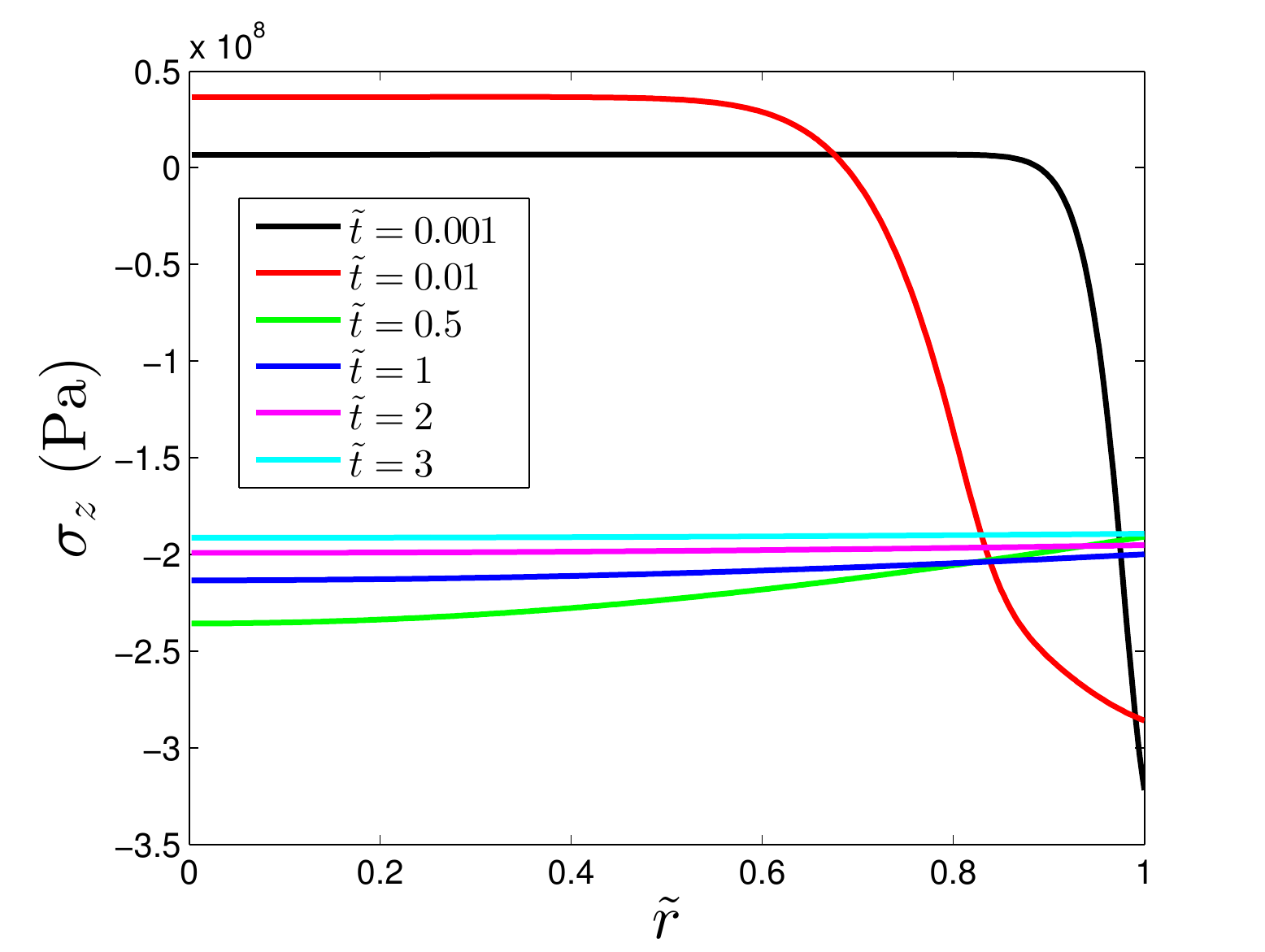}
\caption{}
\end{subfigure} 
\caption{Evolution of stress (represented dimensionally in units of Pa) with time for the case when the electrode particle is constrained in the axial direction. Panels (a), (c), and (e) correspond to a charging condition with (a) $\tilde{J}_0=0.001$, while (b), (d), and (f) to that with $\tilde{J}_0=0.1$. Inset in (e) shows a zoomed-in view at later times.} \label{fig:nointcon_sigma}
\end{figure}

\begin{figure}
\centering
\begin{subfigure}[b]{0.45\textwidth}
\includegraphics[width=\textwidth]{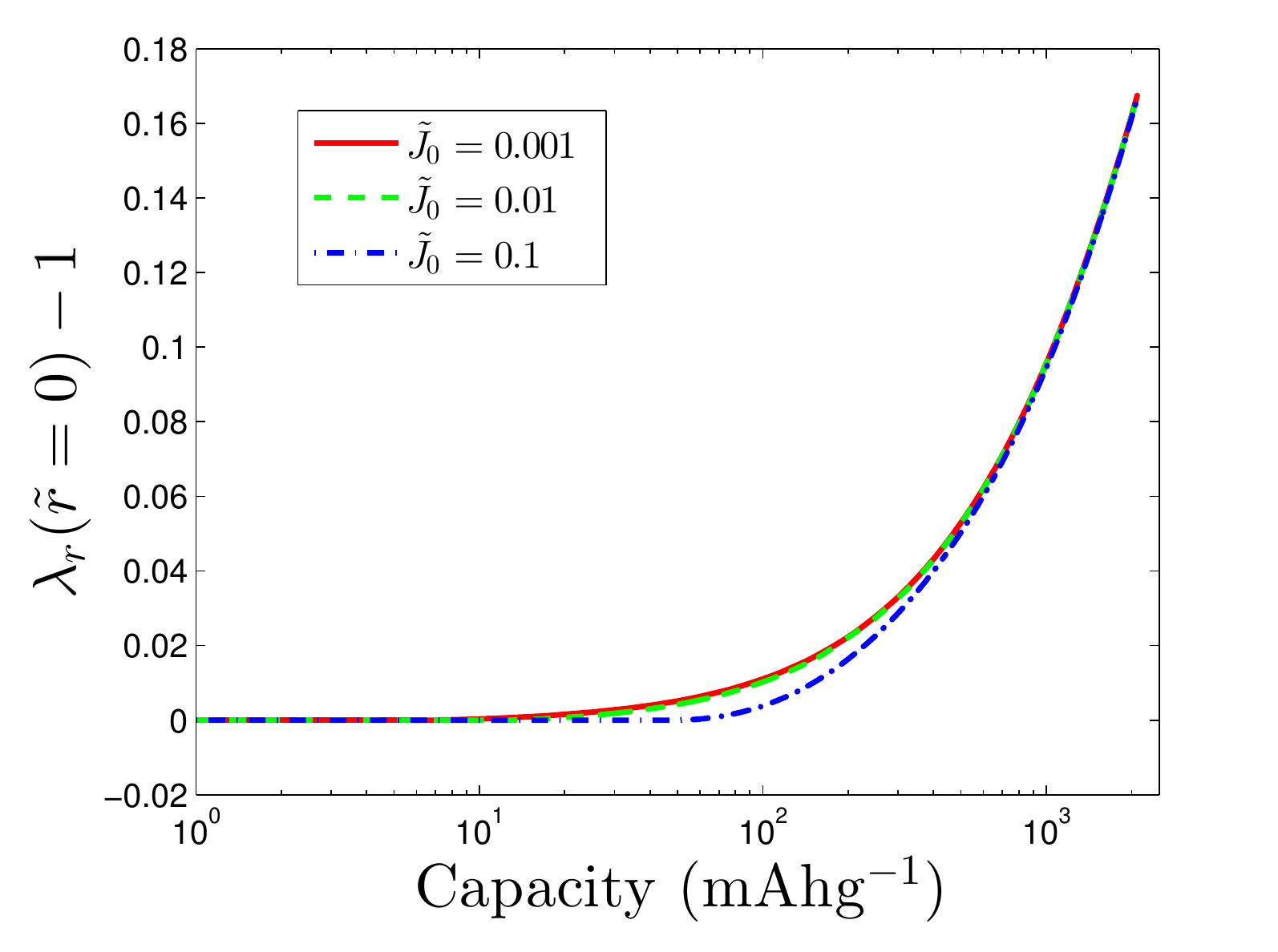}
\caption{}
\end{subfigure}%
\begin{subfigure}[b]{0.45\textwidth}
\includegraphics[width=\textwidth]{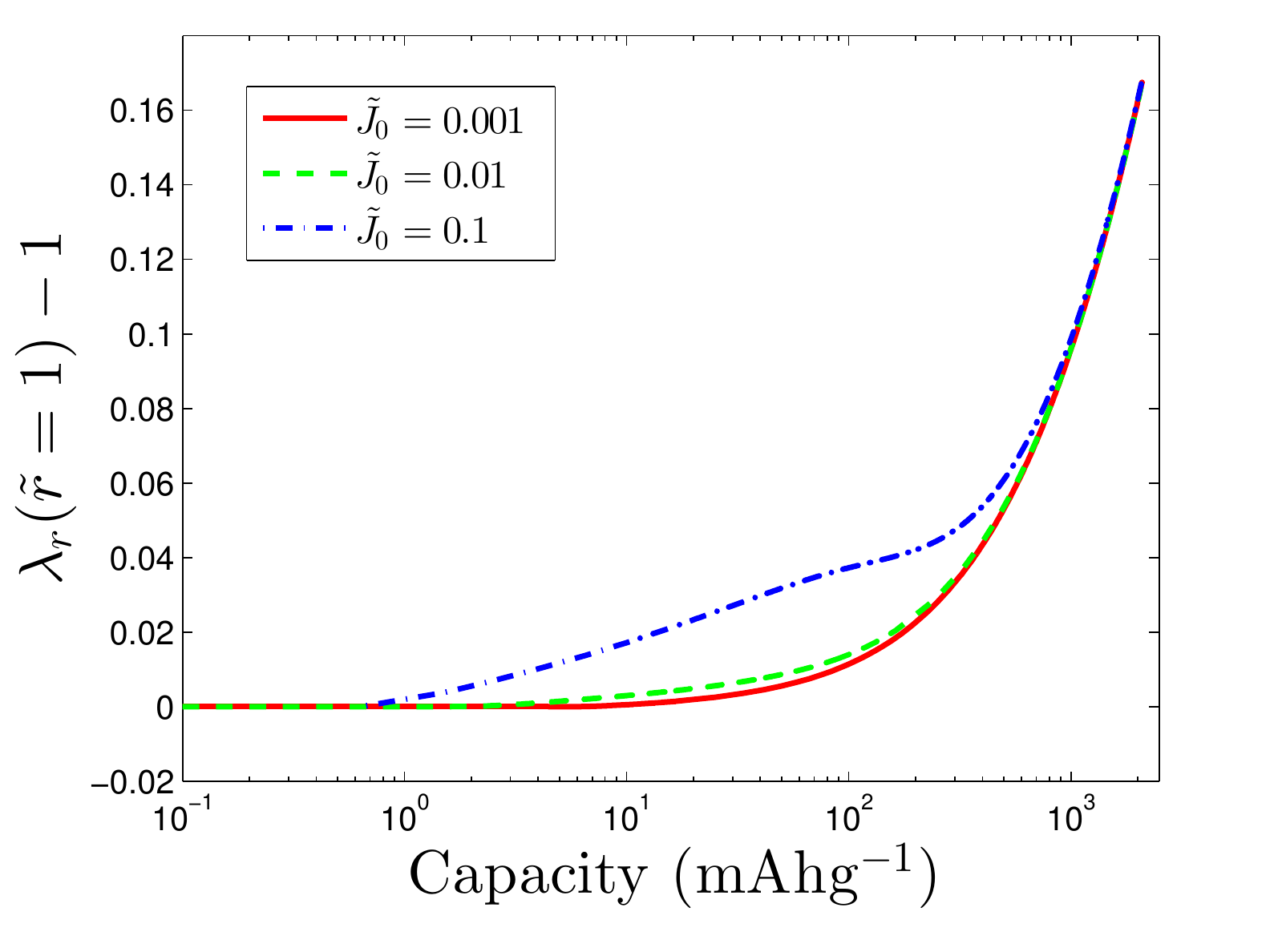}
\caption{}
\end{subfigure}
\begin{subfigure}[b]{0.45\textwidth}
\includegraphics[width=\textwidth]{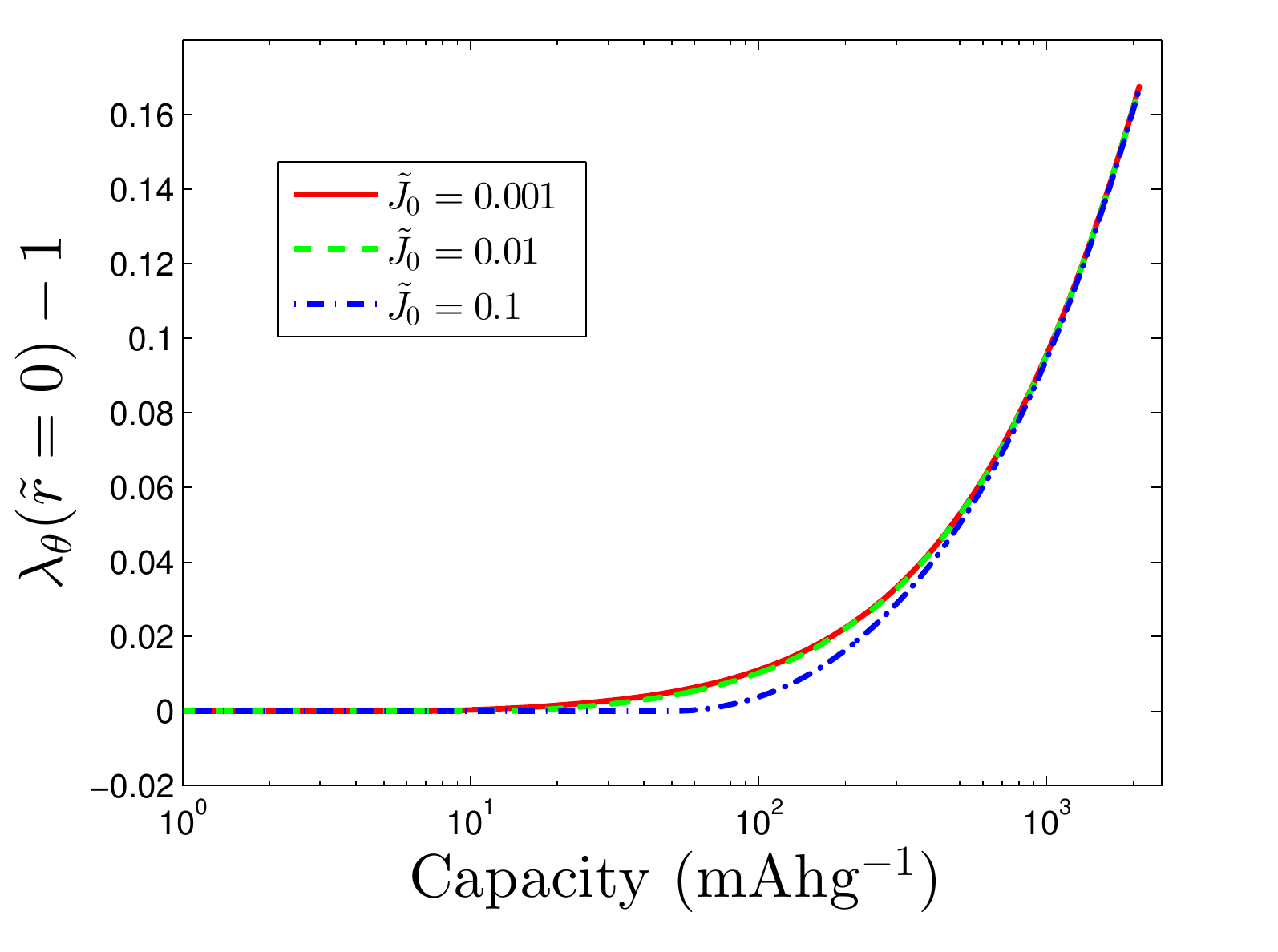}
\caption{}
\end{subfigure}%
\begin{subfigure}[b]{0.45\textwidth}
\includegraphics[width=\textwidth]{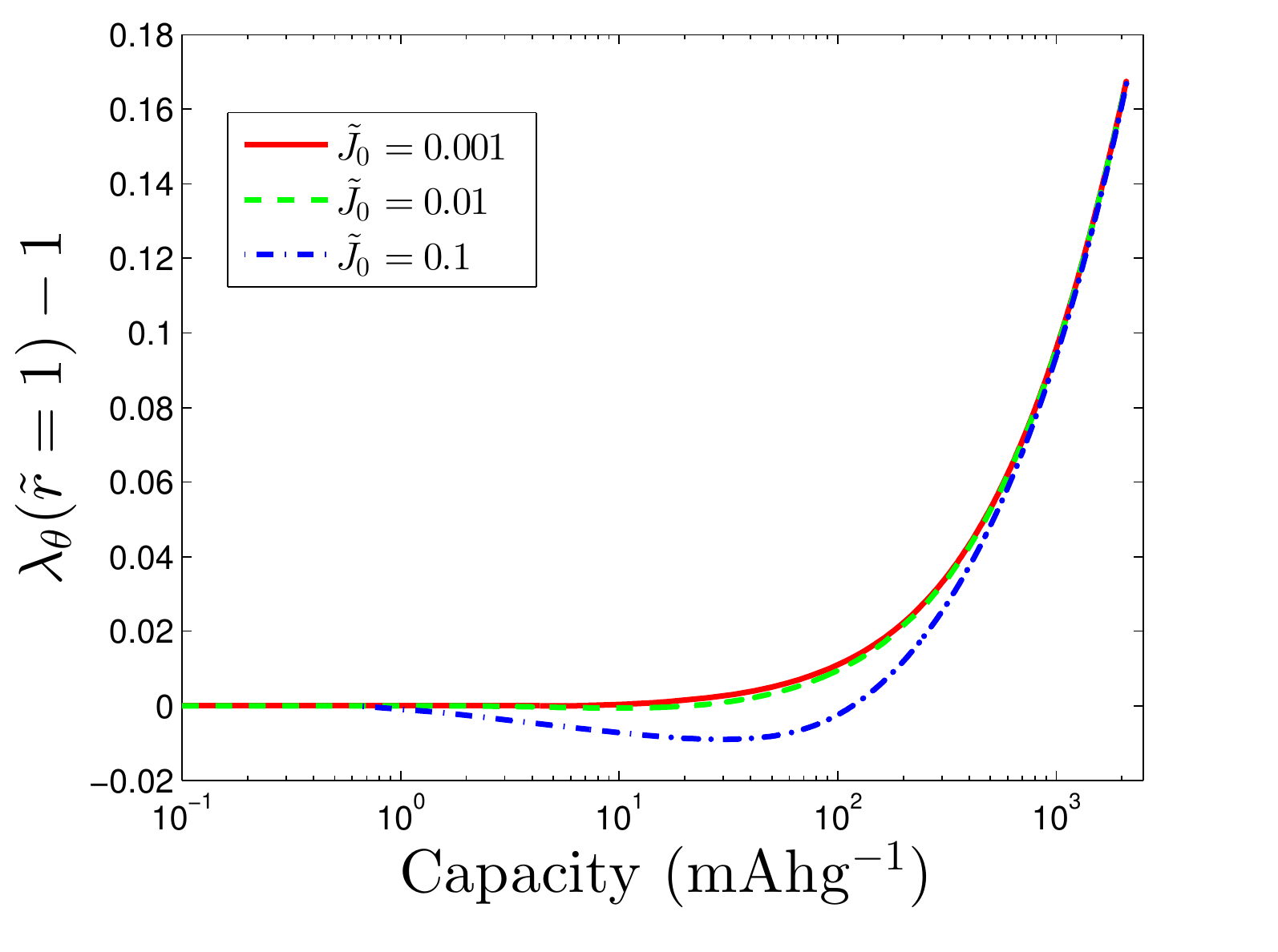}
\caption{}
\end{subfigure}
\begin{subfigure}[b]{0.45\textwidth}
\includegraphics[width=\textwidth]{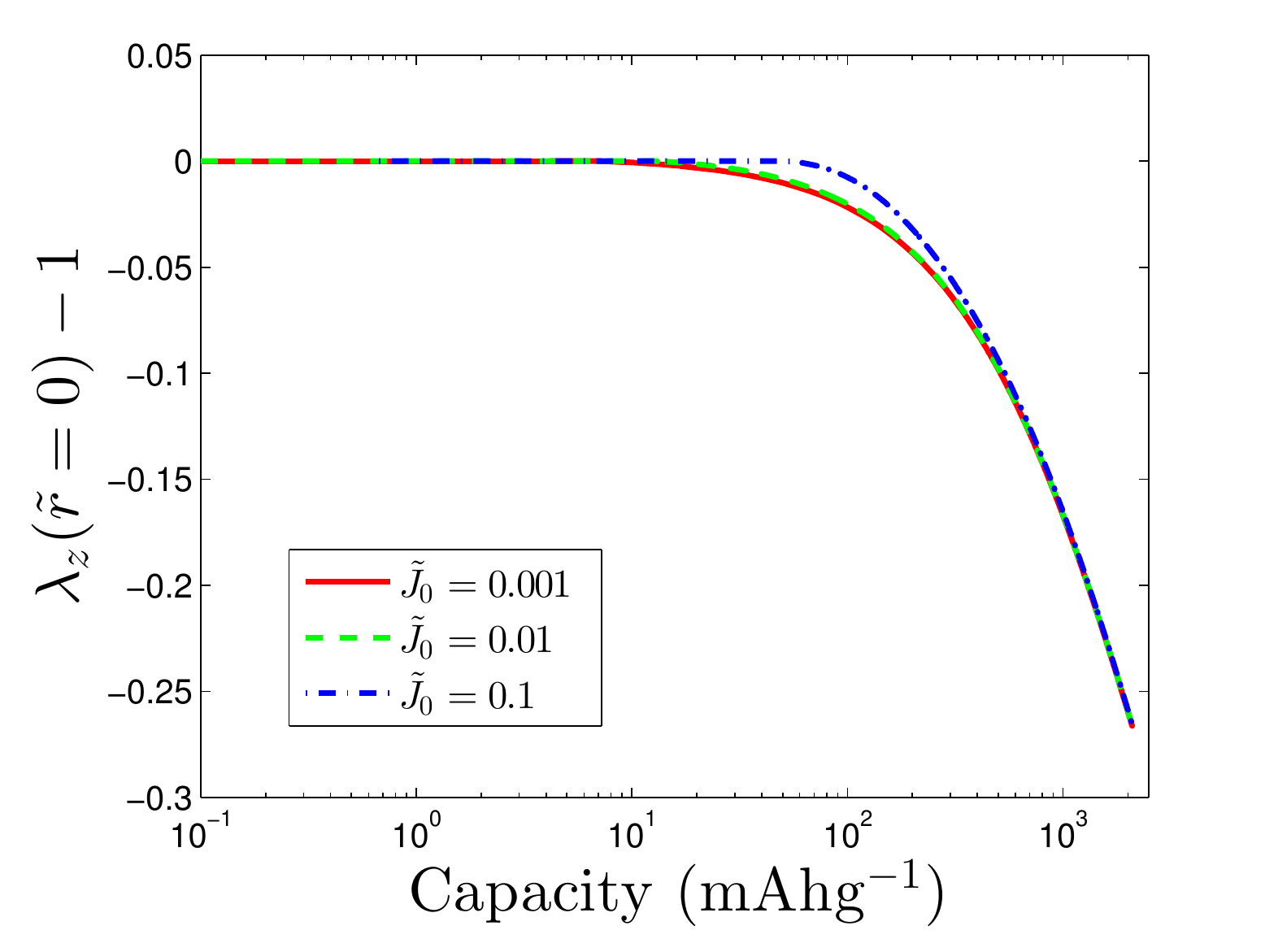}
\caption{}
\end{subfigure}%
\begin{subfigure}[b]{0.45\textwidth}
\includegraphics[width=\textwidth]{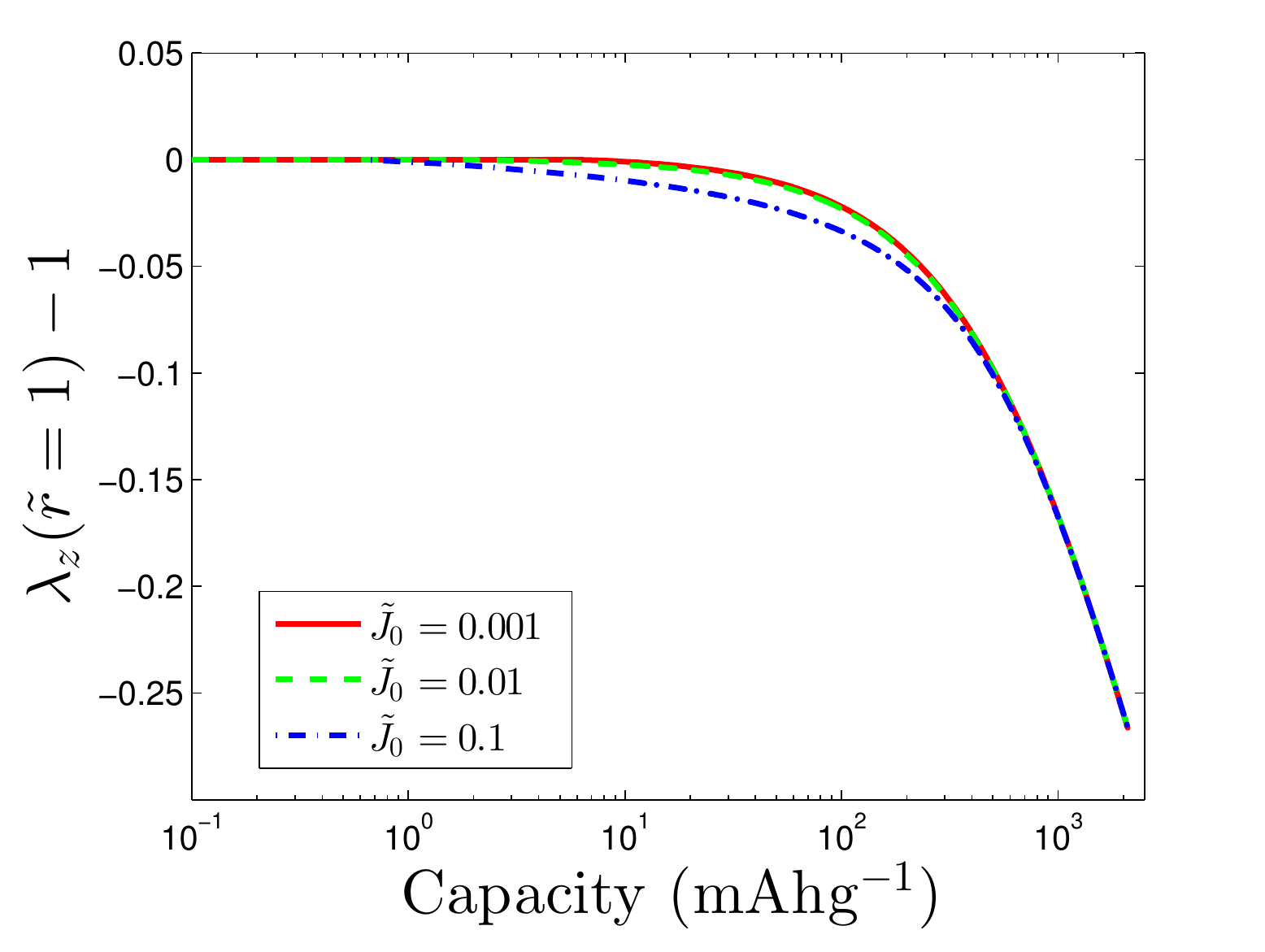}
\caption{}
\end{subfigure} 
\caption{Variation of plastic stretches with increasing lithium concentration represented in terms of the capacity (expressed dimensionally in units of mAhg$^{-1}$) for the case when the particle is constrained in the axial direction. Panels (a), (c), and (e) correspond to the centre ($\tilde{r}=0$), while (b), (d), and (f) to the surface ($\tilde{r}=1$).} \label{fig:nointcon_lambda}
\end{figure}

Fig.~\ref{fig:nointcon_sigma} shows the variation of  stresses  over the radius of the axially constrained cylinder, again corresponding to two different lithiation rates, $\tilde{J}_0=0.001$ and $\tilde{J}_0 = 0.1$. We observe that there is a major difference between the nature of the stress evolution in this case, and that in the unconstrained cylinder case discussed in Sec.~\ref{subsec:generalresults}. Here, the trends observed for $\tilde{J}_0=0.1$ at $\tilde{t}=3$ show up also for $\tilde{J}_0=0.001$ but at later times, $\tilde{t}=30$. But, in the unconstrained case, the trends for $\tilde{J}_0=0.001$ showed no qualitative similarities with the trends for $\tilde{J}_0$ even at $\tilde{t}=300$, as seen in Fig.~\ref{fig:intcon_sigma}. This may be physically explained by the fact that compressive stresses develop quickly in the axial direction due to the end constraint. Furthermore, this constraint necessitates the volumetric expansion (due to lithiation) to be accommodated in the radial and the hoop directions only. This, in turn, leads to a higher increase of stress along both these directions (in distinction with the unconstrained case where there is less inhibition against volumetric expansion). 
We discuss the other important trends in the stress evolution in the context of discussing the evolution of the plastic stretches.

Fig.~\ref{fig:nointcon_lambda} shows the evolution of the plastic stretches again at the centre (as seen in panels (a), (c), and (e)) and at the surface (as seen in panels (b), (d), and (f)) for three values of the influx rate, $\tilde{J}_0=0.001$, $0.01$, and $0.1$. Just as in the unconstrained case, we observe that the plastic stretches do indeed deviate from a value of 1 at the centre, again indicating that the elastic-plastic front moves into the centre.

Note also that the plastic hoop stretch at the surface corresponding to the influx condition $\tilde{J}_0=0.1$ first deviates to a value less than 1, and then, as the concentration builds up, reverses its trend, and increases to a value greater than 1. This indicates that for a relatively high lithiation rate, the surface first undergoes a compressive plastic yielding in the hoop direction followed by a reversal to tensile plastic yielding. This is because at the  surface of the cylinder there is no radial self-constraint. However, in the hoop direction there is a build-up of compressive stresses - leading, ultimately, to a plastic deformation.  With time, the Li diffuses into the centre, and induces deformation there, too. To gain  insight into the nature of this deformation (whether compressive or tensile) it is important to understand first that in an axisymmetric situation both stretches and stresses are equal at the centre for the same capacity values as seen in panels (a) and (c) in Fig.~\ref{fig:nointcon_lambda}. Since $\lambda_r>1$ throughout the cross-section, and over the entire charging time,  $\lambda_\theta$ must necessarily be greater than 1 at the centre, too. Physically, this means that with increasing lithiation, the centre expands, and pushes out against the already plastically deformed outer zone (notably, with $\lambda_\theta<1$). This reverts the stresses in the outer zone from the high compressive values to tensile values -- as can be seen in Fig.~\ref{fig:nointcon_sigma} (d). This, in turn, induces the plastic stretches to invert in nature from $\lambda_\theta<1$ through $\lambda=0$, and ultimately to $\lambda_\theta>1$. It is important to remember, however, that this behaviour is observed only for a relatively high value of lithiation rate (high $\tilde{J}_0$ value). In situations where the lithiation rate is low enough - or, in more general terms, when the relaxation time of lithium diffusion in the silicon is comparable to the rate of lithium influx - such differentiated nature of yielding at the surface and at the centre is not observed, precluding the possibility of observing any reverting trend.

The trend in the evolution of axial stretch can be understood using the same principle. Since the net deformation in the axial direction is physically constrained, the only plastic stretches possible are compressive in nature both at the centre and at the surface throughout the charging time. Again, for a relatively high charging rate, though, plastic yielding at the surface occurs prior to that at the centre while for a low charging rate both the centre and the surface yield almost simultaneously. 

\begin{figure}
\centering
\includegraphics[width=0.5\textwidth]{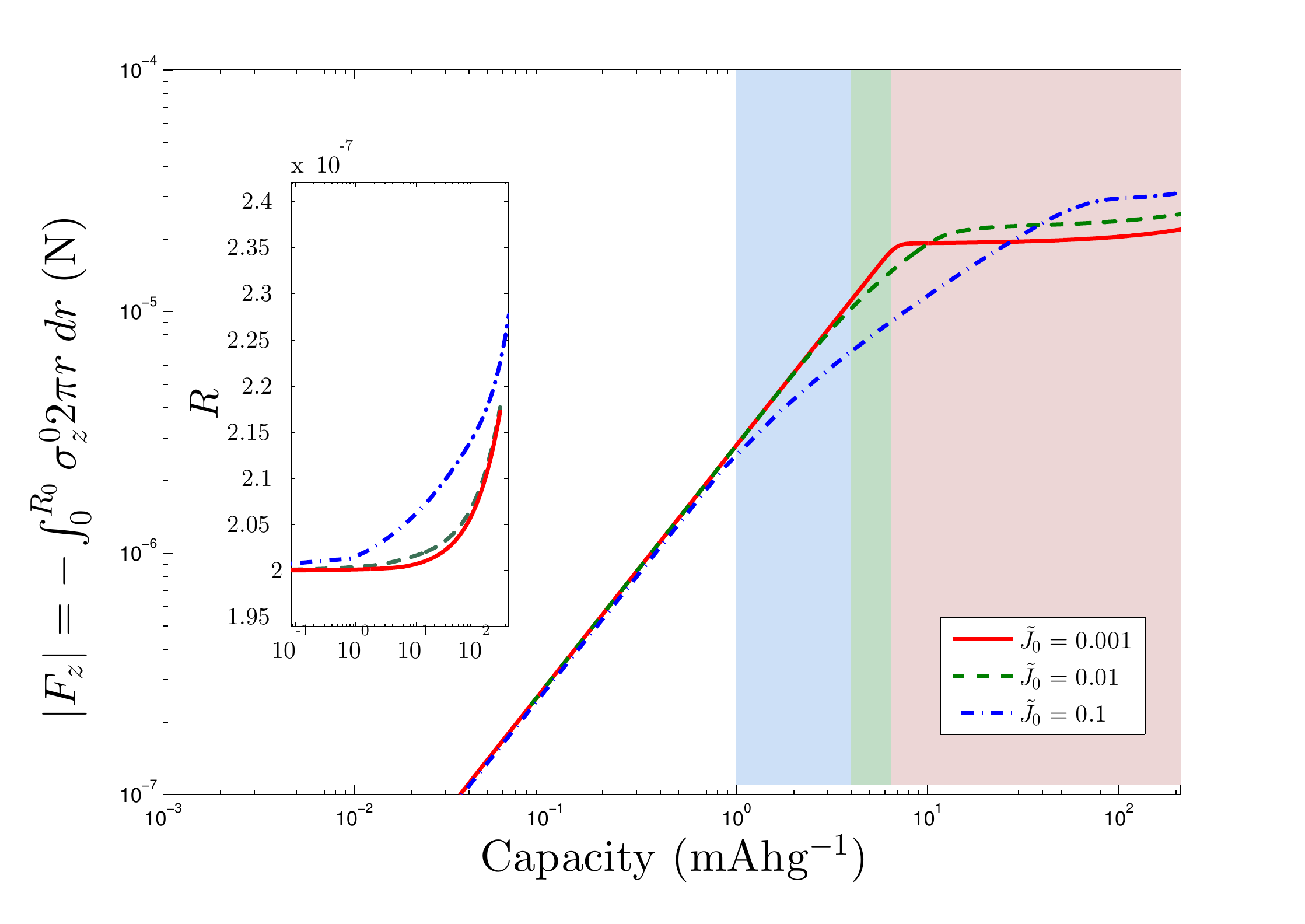}
\caption{Variation of the compressive force (expressed dimensionally in units of N) in the axial direction with increasing lithium concentration represented in terms of the capacity (expressed dimensionally in units of mAhg$^{-1}$). The shaded regions indicate the zones where plastic deformation is present (anywhere within the cross-section) corresponding to each influx rate. Inset shows the evolution of the radius with increasing lithium concentration (again expressed in terms of the capacity) for the three different influx rates.} \label{fig:Fz}
\end{figure}

Fig.~\ref{fig:Fz} shows the variation of the compressive axial force (in units of N) as lithium concentration (expressed in terms of the capacity) builds up in the anode for three different values of the influx rate. Each of the three shaded regions: blue, green, and pink - the green region superimposed on the blue, and the pink on the green - denotes the zone where plastic deformation has set in corresponding, respectively, to the three different values of the lithiation rates, $\tilde{J}_0=0.1$, $0.01$, and $0.001$. The concentration value at which the plastic deformation sets in for each influx rate is obtained by comparison with Fig.~\ref{fig:nointcon_lambda} (f) (since plastic deformation sets in first at the surface for higher value influx rate, and almost simultaneously at the centre and at the surface for lower influx rate). As expected, the plastic deformation starts at a relatively lower value of the concentration build-up for a higher value of the lithiation rate because the fast constitutive changes brought about through a higher rate of influx induces higher stresses. As soon as the material begins to flow plastically, however, the ``growth" rate of the stresses drops (this is clearly seen in Fig.~\ref{fig:nointcon_sigma} (e) and (f)) so that for a given change in concentration, a proportionately lower increase in the magnitude of the compressive axial force is manifested. The axial force, nevertheless, continues to increase, albeit at a lower rate, due to the contribution of the increasing value of the radius. In connection with this, it is important to recall that the  the axial force is computed from the first Piola-Kirchhoff stresses which denotes the stresses in the current configuration with respect to the reference configuration, and consistently with that, the reference radius is used. However, Fig.~\ref{fig:nointcon_sigma} shows the Cauchy stresses, and, hence, a discussion of the axial force evolution requires the  consideration of the stress and the changing radius. This may be explicitly seen in the inset of Fig.~\ref{fig:Fz} which shows that the radius does, in fact, start to increase for each value of the influx rate corresponding to those values of the concentration where the plastic deformation begins (as represented in the main figure through the shaded regions). Since the radius does not deviate much from the reference value when $\tilde{J}_0=0.001$, and $\sigma_z$ has also relaxed to a constant value (essentially determined by the yield stress value), the axial force is practically constant following plastic yield. For $\tilde{J}_0=0.1$, on the other hand, even though $\sigma_z$ relaxes to a practically constant value again, the radius keeps increasing significantly with lithiation, leading, in turn, to an increase in axial force.  

\subsection{Buckling}

\begin{figure}[t!]
\begin{subfigure}[b]{0.49\textwidth}
\includegraphics[width=\textwidth]{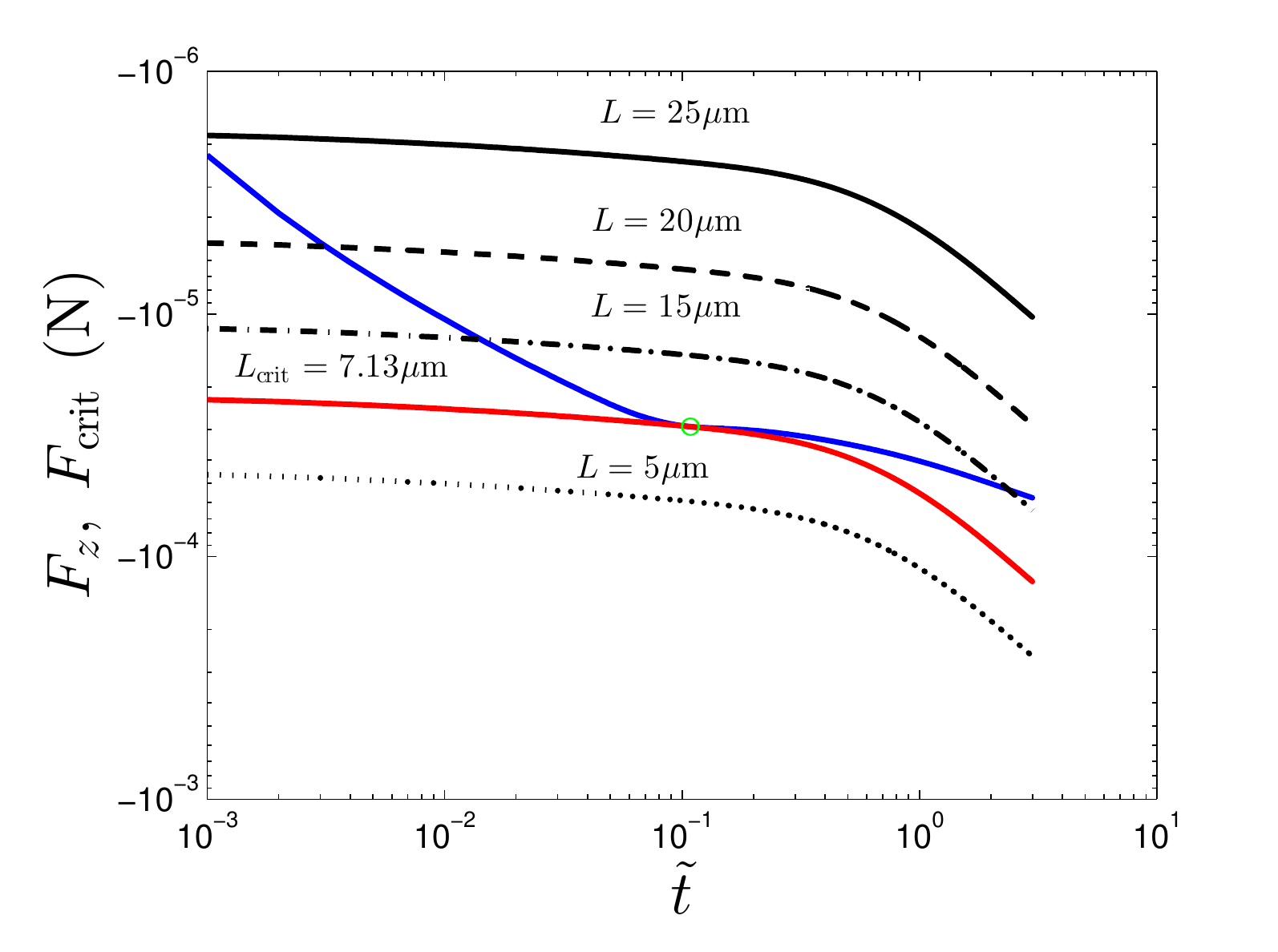}
\caption{}
\end{subfigure}%
\begin{subfigure}[b]{0.45\textwidth}
\includegraphics[width=\textwidth]{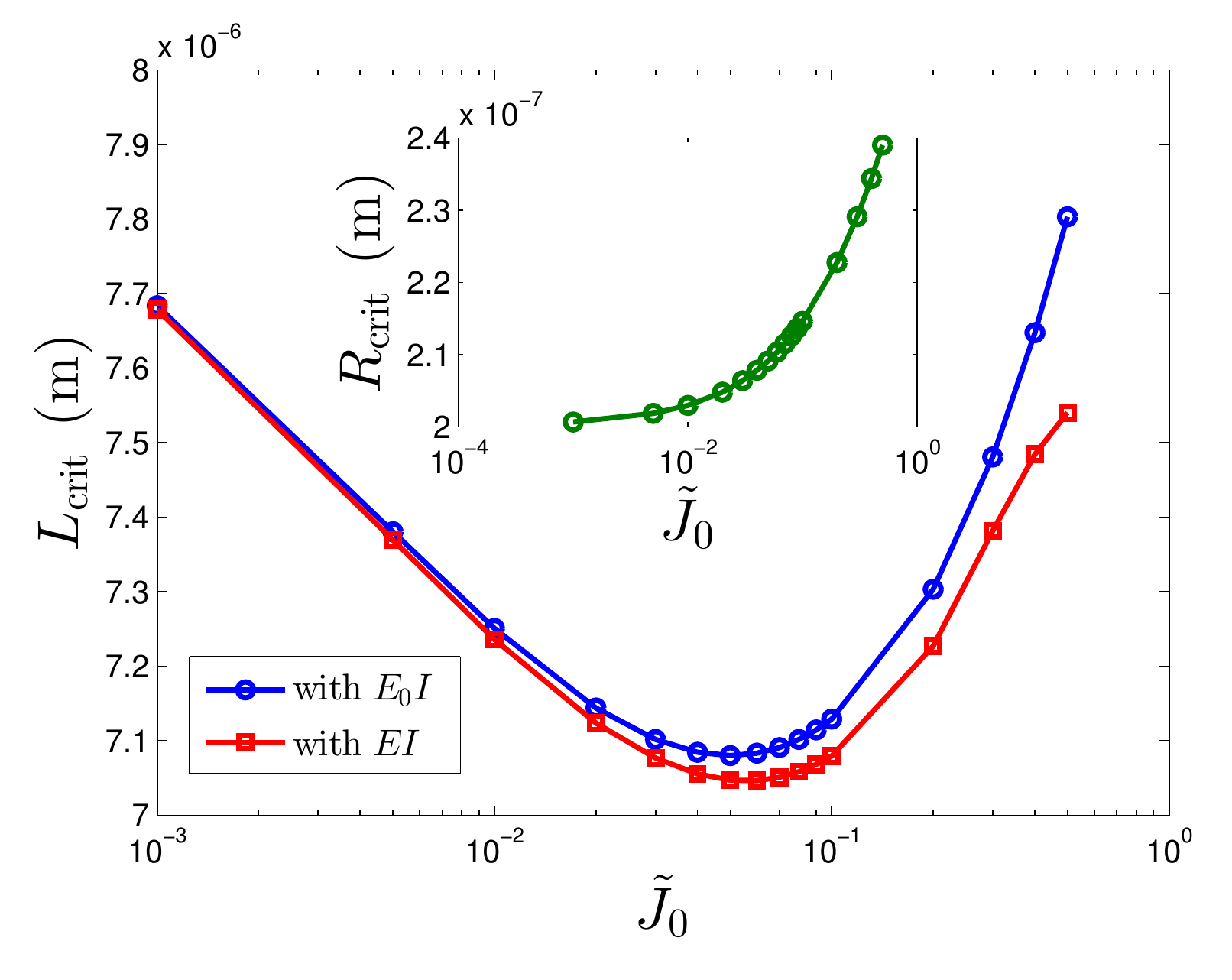}
\caption{}
\end{subfigure}
\caption{Panel (a) shows a representative set of load curves of the cylinders corresponding to four different values of the cylinder length including the critical one (the red curve) for which the load curve just touches the axial force curve when the influx rate $\tilde{J}_0=0.1$ based on the first modification (see text for details) of the classical Euler buckling criterion. Panel (b) shows the variation of the critical length of the cylindrical anode (expressed dimensionally in units of m)  below which buckling never occurs with the rate of lithiation in terms of the non-dimensional flux. Critical lengths calculated on the basis of both modifications of the Euler buckling criterion are shown. Inset of right panel shows the variation of the final radius of the cylindrical anode again with the rate of lithiation at the critical point when buckling occurs.} \label{fig:Lcr}
\end{figure}

We are now in a position to discuss the role played by the axial force in the buckling of the cylinder. In Fig.~\ref{fig:Lcr}(a) we show the changing axial force $F_z$ as a function of time for an influx rate $\tilde{J}_0=0.1$ (blue curve). Also shown is the critical axial load for buckling $F_{\rm{crit}}$ (which varies with time as the cylinder radius increases) for different cylinder lengths $L$ (for clarity we show the plots corresponding to the first modified criterion only). For a given length $L$ the intersection of these two curves gives the time at which buckling will occur. We find that for short enough lengths buckling will never occur - the increase in radius is more than enough to compensate for the increase in axial stress (see dotted curve for $L=5~\mu$m). Thus for a given influx rate, there is a critical length, $L_{\rm{crit}}$ below which the cylinder will not buckle. 

Next, we evaluate the critical length, numerically, on the basis of the two different modifications of the classical Euler buckling criteria as discussed in Sec.~\ref{subsec:special}, and corresponding to different values of the influx rate. Note that if we do not consider these modifications, the load curve for a particular electrode particle length is a horizontal line, and in that situation, no critical value of length can be found. Fig.~\ref{fig:Lcr}(b) shows the critical length for influx rates spread across more than two orders of magnitude, and for both modifications of the classical buckling criterion. The blue curve (with circular markers) corresponds to the first modification while the red curve (with square markers) corresponds to the second modification. 

We observe similar qualitative trends in the critical lengths for both the modifications. When $\tilde{J}_0 \lesssim 0.05$, the critical length decreases with increasing influx rate, while for higher values of $\tilde{J}_0$, the critical length increases. In other words, when $\tilde{J}_0 \lesssim 0.05$ an electrode particle of given length is increasingly more susceptible to buckling with increase in the lithiation rate, and that the behaviour is reversed for higher lithiation rates. This can be physically understood on the basis of the interplay between the stress build-up and the increasing radius. It is intuitive to understand that higher stresses in the axial direction leads to higher axial compressive forces. On the other hand, increasing radius makes the electrode particle less slender -- equivalently, more ``stubby" -- and, hence, less prone to buckling. As discussed previously, the concentration gradients become sharper with increase in the influx rate, and this, in turn, induces higher stresses. For lithiation rates up to $\tilde{J}_0=0.05$, however, the growth in the radius for a given charging time is significantly less than that for higher lithiation rates; this is clearly seen in the inset of Fig.~\ref{fig:Lcr}~(b). Therefore, for $\tilde{J}_0 \lesssim 0.05$, it is the influence of the stresses which predominates over that of the increasing radius so that the critical length decreases; for higher values of $\tilde{J}_0$, the stabilising influence of the increasing radius starts to dominate -- allowing for increasing critical lengths up to which the cylindrical electrode does not buckle. 

We also observe that for each influx rate, the critical lengths corresponding to the second modification are lower than those corresponding to the first modification. This may be explained as follows. An increase in the influx rate results in higher build-up of Li concentration for a given charging time, and leads to an increase in the value of the radius, and makes the cylinder stubbier by increasing the value of the flexural rigidity. In the same way, since the value of the rate of the change of elastic modulus with concentration ($\eta_E$) is negative, a higher concentration build-up resulting from a higher influx rate leads to a decrease in the value of the elastic modulus which decreases the flexural rigidity, and makes the cylinder more prone to buckling. Thus, for a given value of the influx rate, the use of the second modification of the buckling criterion which incorporates the concentration dependence of the elastic modulus results in a lower value of the critical length.

However, it is important to note that since the nature of the plots for both modifications is qualitatively the same, the influence of the increasing radius in determining the critical behaviour is more important than that of the decreasing elastic modulus with concentration. Therefore, in further discussions, we use only the first modification to obtain further insight into critical buckling behaviour. 

Before proceeding, however, we observe that Fig.~\ref{fig:Lcr}~(b) shows that the critical length can be taken as 7 $\mu$m for all values of the influx rate corresponding, of course, to the values of the material properties and parameters listed in Table~\ref{table:values}. From a design perspective, therefore, it would be interesting to see if an estimate of such a critical length can be made from a simple calculation. Towards that end, we note that for buckling to occur, the axial compressive force must match the critical buckling force. We choose the yield stress, $\sigma_f$, as the representative value that determines the compressive force. Further considering no influence from the change in radius, we have:
\begin{align}
\sigma_f \pi R_0^2 &= \frac{\pi^2 E_0}{L^2} \left(\frac{\pi R_0^4}{4}\right),  \\
\text{or} \quad L &= \frac{1}{2}\sqrt{\frac{E_0}{\sigma_f}} \pi R_0,
\end{align}
which gives $L \sim 8 \mu\rm{m}$ which matches quite closely with the value determined from the full model. 

\begin{figure}[ht!]
\centering
\includegraphics[width=0.5\textwidth]{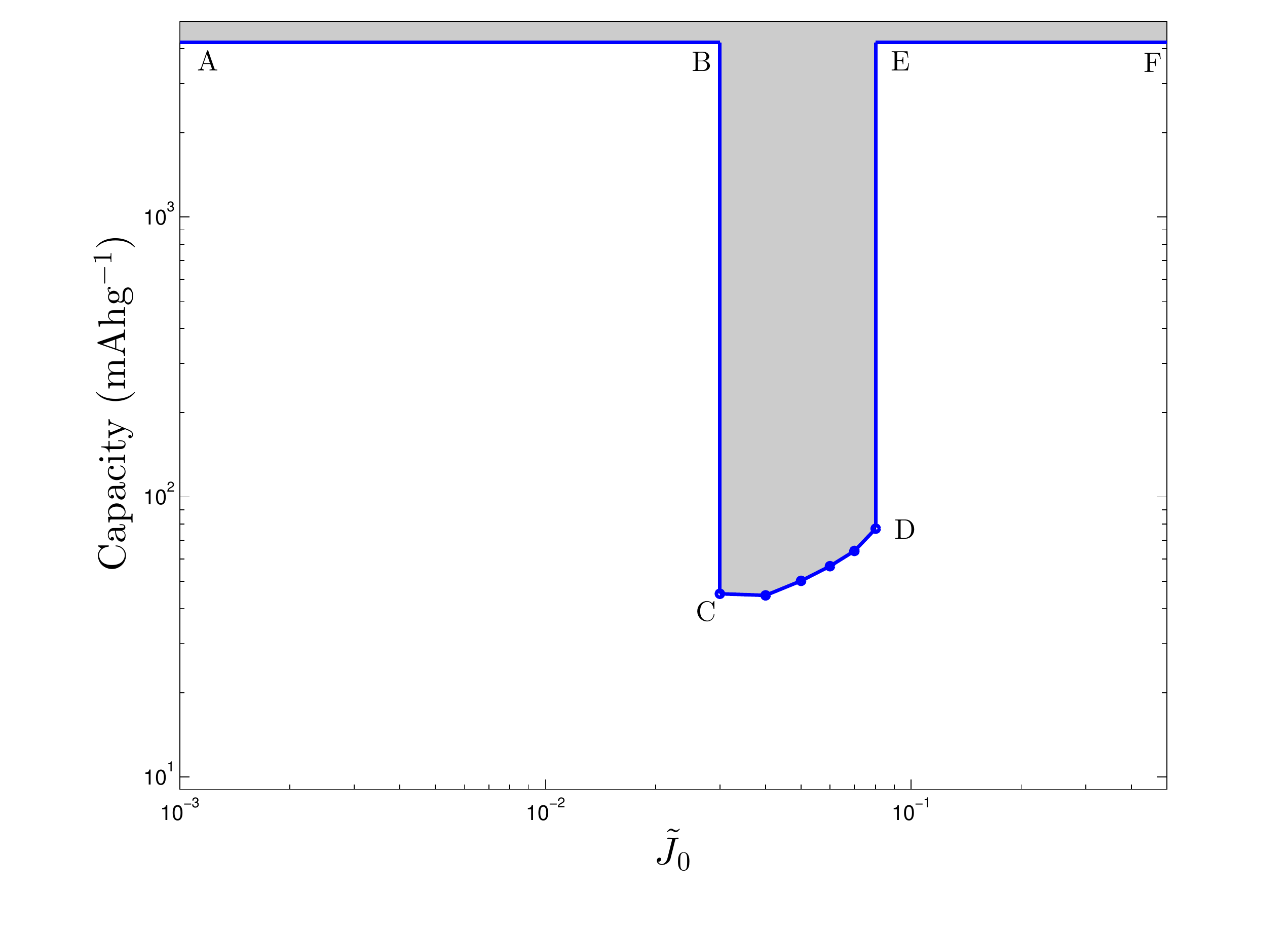}
\caption{The unshaded region bounded by ABCDEF denotes situations for which a cylinder of a given length will not buckle. C and D denote situations where this length is also the critical length for two different values of the influx rate. AB and EF denote the critical situation corresponding to full lithation.} \label{fig:Capexp}
\end{figure}

In Fig.~\ref{fig:Capexp} the curve CD denotes the critical capacity values corresponding to different influx rates at which a cylinder of a given length will buckle (using the first modification of the Euler buckling criterion). The unshaded therefore region represents the amounts of lithiation (capacity values) that are possible without buckling for different influx rates: For influx rates between C \& D, the cylinder does not buckle as long as the amount of lithiation is within the unshaded region. For influx rates to the left of C and to the right of D, although the cylinder does not ever buckle, the amount of lithiation has an upper bound of 4200 mAhg$^{-1}$, represented by the horizontal lines AB and EF. The points C and D denote situations where this length is also the critical length (refer Fig.~\ref{fig:Lcr}) for the values of the influx rates corresponding to them. Hereafter, we will refer to points such as C and D as kink points.   

\begin{figure}[ht!]
\centering
\includegraphics[width=0.5\textwidth]{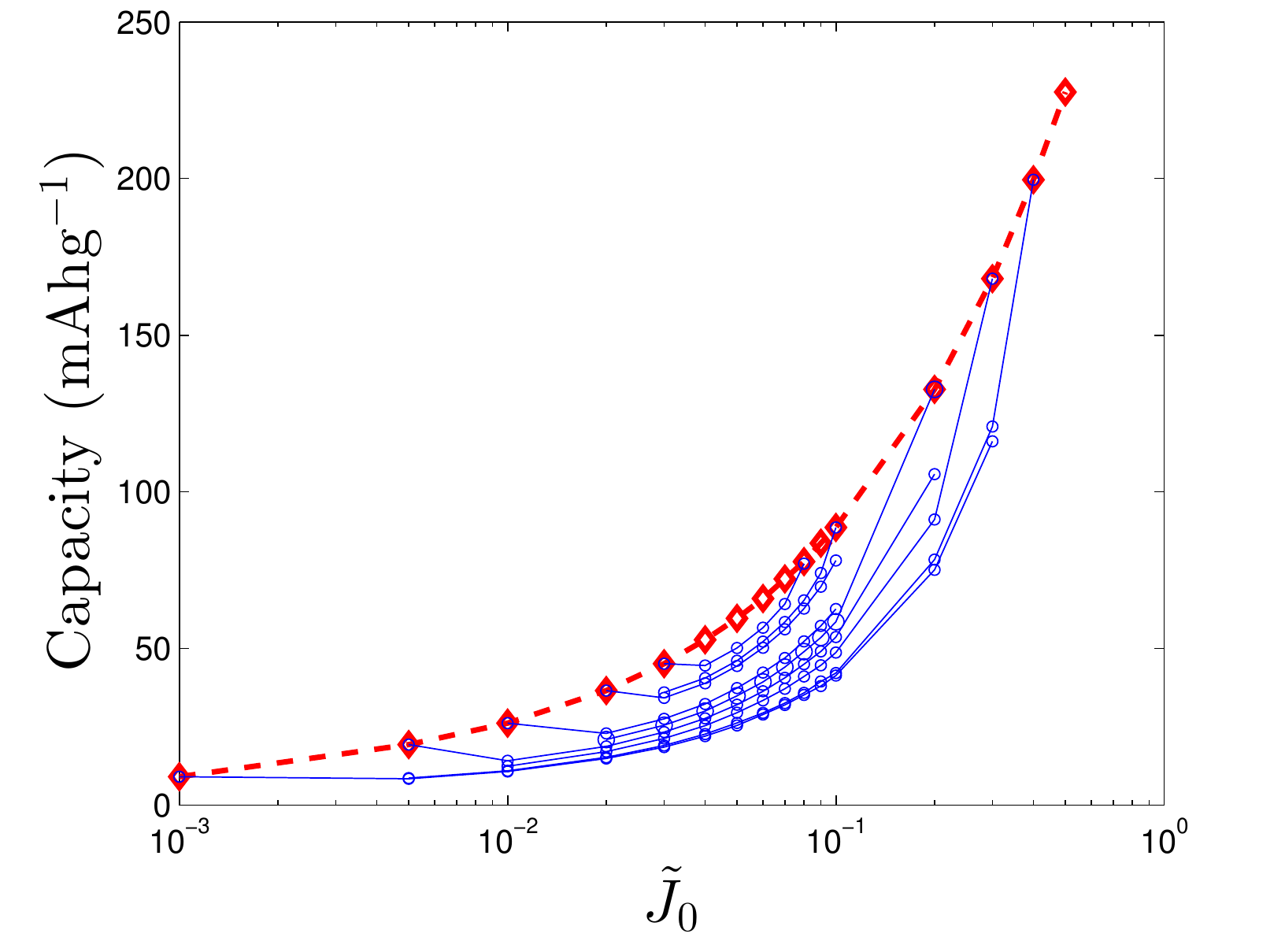}
\caption{The red, dashed plot shows the values of the capacity corresponding to critical values of the length at which the cylinder buckles. Each blue, bold line is a locus of the critical capacity values at which the cylinder of a given length buckles.} \label{fig:Capint}
\end{figure}

In Fig.~\ref{fig:Capint}, we plot (red, dashed) the capacity values at which the cylinder buckles for different values of the influx rate corresponding to the critical lengths as discussed in Fig.~\ref{fig:Lcr}(b). We call this plot the capacity envelope. Additionally, we also show the locii (blue, bold) of the critical values of the capacity when the cylinder of a given length buckles. In general, this length will not be the critical one. However, when this length does coincide with the critical length at a particular influx rate, the locus (blue, bold) meets the capacity envelope (red, dashed) at a kink point (similar to points C and D in Fig.~\ref{fig:Capexp}). Thus, the capacity envelope is also a plot of the kink points for different given lengths. Furthermore, referring to Fig.~\ref{fig:Lcr}, we note that the difference in the influx rate values at the kink points corresponding to a particular locus is an indicator of the cylinder length. The closer the kink points are, the smaller is the length. In other words, the length decreases as the locii approach the capacity envelope.

The increasing trend in the values of the capacity -- both the values corresponding to the capacity envelope as well as those belonging to the different locii -- may be physically interpreted as follows. Although a higher value of the influx rate may quickly result in sharper gradients in the Li concentration leading, in turn, to higher stresses giving the impression that buckling might occur for lower values of the capacity when the influx rate is higher, there are two mitigating factors. First, a higher influx rate while it creates sharp concentration gradients quickly, also ``pumps" in more Li in that short duration so that the absolute value of the capacity is indeed high when the cylinder buckles. Second, as discussed extensively in the preceding paragraphs, the increase in the value of the Li concentration is accompanied by a growth in the radius which makes the cylinder ``stubbier" so that higher amounts of Li need to be inserted before the compressive axial force transgresses the buckling load. In this connection, it is very important to note that the particular form of the buckling criterion used with its dependence on the current radius means that the load curve itself changes to accommodate higher compressive axial forces before buckling sets in. This is clearly seen in panel (a) of Fig.~\ref{fig:Lcr}.

\section{Conclusions}

We have presented a framework to model the behaviour of a cylindrical Si anode particle as it undergoes lithiation during charging. First, we have applied this model to a situation where no external physical constraint is imposed on the cylindrical particle, and then to a situation where the cylinder is constrained axially. Observing the trends in the evolution of the stresses and the plastic stretches, we find that there are important differences between the plastic stretches in the present cylindrical case and those in the spherical case investigated earlier by \citet{2012JMechPhysSolidsCui}. The plastic stretches \emph{do} deviate from 1 even at the centre here due to the absence of a hydrostatic state of stress that is present in the spherical case. Even between the unconstrained and the constrained case, significant qualitative differences are observed in the nature of the stress evolution. In the unconstrained case, the stresses corresponding to a lower influx rate ($\tilde{J}_0=0.001$) show no similarities in spatial variation over the cylinder cross-section with those corresponding to a higher influx rate ($\tilde{J}_0=0.1$) even after a long charging time. In contrast to this, in the constrained case, the stresses corresponding to the lower influx rate start showing qualitative similarities in their spatial variation with those corresponding to the higher influx rate after a sufficiently long charging time. Furthermore, in the constrained case, plastic deformation sets in at a lower value of concentration for higher influx rate compared to the lower influx rate. This induces a change in the growth rate of the axial forces. For lower influx rates, the axial forces plateau off, while for higher influx rates, the axial forces continue to grow due to the influence of the growing radius -- an influence that is significantly lower at lower influx rates. 

We have also presented a simple framework to examine the critical lengths at which the constrained cylindrical particle may buckle because of the growing axial forces. This framework includes two criteria for finding the critical values, both based on modifications of the classical Euler buckling formula. The first modification takes into account the sole influence of the growing radius while the second modification takes into account the influence of the decreasing value of the modulus of elasticity in addition to that of the radius. Using these criteria, we find that the influence of the growing radius is stronger than that of the changing modulus. We observe that the critical lengths first decrease up to a certain value of the influx rate, and then start to increase indicating that the propensity to buckle is highest for the mid-values of the influx rates. The reversal in trend of the critical lengths is explained on the basis of the growing radius which while it increases the axial forces also makes the cylinder stubbier, and hence less prone to buckling. This argument is further reinforced by the increasing trend in the capacity values at buckling for a cylinder of a given length with increasing influx rate despite the fact that higher influx rates may lead to higher stresses through sharper gradients in the concentration. For design considerations, we find that the critical buckling length is not influenced strongly by the influx rate. Further, a simple estimate of the critical length can be obtained by using the formula $L = \frac{1}{2} \left( E_0/\sigma_f \right)^{1/2} \pi R_0$.

\section*{Acknowledgements}
J.C., S.J.C., and A.G.  acknowledge support from the EPSRC through Grant No. EP/I017070/1. C.P.P. acknowledges support from the EPSRC through Grant No. EP/I01702X/1. A.G. is a Wolfson/Royal Society Merit Award Holder and acknowledges support from a Reintegration Grant under EC Framework VII. J.C. thanks Prof. Allan Bower (Brown University), Dr Giovanna Bucci (MIT), and Prof. Jianmin Qu (Northwestern University) for clarifications in their papers.

\bibliographystyle{chicago}
\bibliography{refs}



\end{document}